\documentclass[prd,preprint,superscriptaddress,nofootinbib,longbibliography]{revtex4}
\usepackage{comment}
\usepackage{graphicx}
\usepackage{dcolumn}
\usepackage{bm,amsfonts,amsthm,amsmath,amssymb,array}
\usepackage{bbm}
\usepackage[]{epsfig,graphics}
\usepackage{comment}
\usepackage[T1]{fontenc}
\usepackage{lipsum}
\usepackage[utf8]{inputenc}
\usepackage{ulem}
\usepackage{multirow}
\usepackage{color}
\usepackage{lastpage}
\usepackage[sort&compress]{natbib}
\usepackage{enumerate}
\usepackage{wasysym}
\usepackage{relsize}
\usepackage{physics}
\usepackage{mathrsfs}
\usepackage{tensor}
\usepackage{xfrac}
\usepackage{siunitx}
\usepackage{setspace}  
  \usepackage[
         colorlinks=true,     
         urlcolor=blue,       
         ]{hyperref}
\usepackage{caption}
\usepackage{subcaption}

\hypersetup{
    unicode=false,          
    pdftoolbar=true,        
    pdfmenubar=true,        
    pdffitwindow=false,     
    pdfstartview={FitH},    
    pdftitle={My title},    
    pdfauthor={Author},     
    pdfsubject={Subject},   
    pdfcreator={Creator},   
    pdfproducer={Producer}, 
    pdfkeywords={keyword1} {key2} {key3}, 
    pdfnewwindow=true,      
    colorlinks=false,       
    linkcolor=red,          
    citecolor=green,        
    filecolor=magenta,      
    urlcolor=cyan           
}

\newenvironment{itemize*}
  {\begin{itemize}
    \setlength{\itemsep}{0pt}
    \setlength{\parskip}{0pt}}
  {\end{itemize}}

\newenvironment{enumerate*}
  {\begin{enumerate}
    \setlength{\itemsep}{0pt}
    \setlength{\parskip}{0pt}}
  {\end{enumerate}}

\newenvironment{description*}
  {\begin{description}
    \setlength{\itemsep}{0pt}
    \setlength{\parskip}{0pt}}
  {\end{description}}

\def\ben{\begin{enumerate*}}
\def\een{\end{enumerate*}}
\def\bi{\begin{itemize*}}
\def\ei{\end{itemize*}}
\def\bd{\begin{description*}}
\def\ed{\end{description*}}
\def\be{\begin{equation}}
\def\ee{\end{equation}}
\def\bea{\begin{eqnarray}}
\def\eea{\end{eqnarray}}
\def\bfl{\begin{flushleft}}
\def\efl{\end{flushleft}}
\def\bigskip{\;\;\;\;\;\;\;}
\def\over{\begin{flushleft}}

\usepackage{xcolor}

\def\ben{\begin{enumerate*}}
\def\een{\end{enumerate*}}
\def\bi{\begin{itemize*}}
\def\ei{\end{itemize*}}
\def\bd{\begin{description*}}
\def\ed{\end{description*}}
\def\be{\begin{equation}}
\def\ee{\end{equation}}
\def\bfl{\begin{flushleft}}
\def\efl{\end{flushleft}}

\newcommand{\Hcon}{\mathcal{H}}

\newcommand{\LM}{Lemaître}

\newcommand{\half}{\frac{1}{2}}

\newcommand{\pd}{\dot{\phi}}

\newcommand{\vdd}{\ddot{\varphi}}
\newcommand{\ld}{\dot{\lambda}}
\newcommand{\ldd}{\ddot{\lambda}}
\newcommand{\lds}{\dot{\lambda}^2}

\newcommand{\vp}{{\varphi}^{\prime }}
\newcommand{\vpd}{\dot{\varphi}}
\newcommand{\vps}{{\varphi}^{\prime \; 2}}
\newcommand{\vpp}{{\varphi}^{\prime \prime}}
\newcommand{\lp}{{\lambda}^{\prime}}
\newcommand{\lps}{{\lambda}^{\prime \; 2}}
\newcommand{\lpp}{{\lambda}^{\prime \prime}}

\newcommand{\dlm}{\Delta_\phi {\cal L}_m}

\begin{document}

\title{Waiting for Inflation: A New Initial State for the Universe}

\author{Brandon Melcher} 
\email{brandon@appliedphysics.org}
\affiliation{Applied Physics, PBC 477 Madison Avenue, New York, NY 10022}

\author{Arnab Pradhan} \email{arpradha@syr.edu} 
\affiliation{Department of Physics, Syracuse
  University, Syracuse, NY 13244, USA}

\author{Scott Watson}
\email{gswatson@syr.edu}
\affiliation{Department of Physics, Syracuse
  University, Syracuse, NY 13244, USA}

\date{\today}

\begin{abstract}
We propose a cosmological lingering phase for the initial state prior to inflation which would help address the singularity problem of inflation. The universe begins with a constant (Hagedorn) temperature and then transitions into an inflationary universe while preserving the Null Energy Condition (NEC). In such a universe time is presumably emergent, calling the age of the universe into question.
We first consider the phase space of positive spatial curvature models within General Relativity and with matter sources that respect the NEC.
Depending on the duration of the post-lingering inflation these models can produce a small amount of observable spatial curvature in the Cosmic Microwave Background.
We also discuss how lingering can arise with or without spatial curvature in theories of quantum gravity when considering the thermodynamic scaling of particles and its impact on the early universe. The string theory dilaton is essential to the dynamics. There are many open questions that remain.
\end{abstract}

\maketitle
\thispagestyle{empty}
\tableofcontents
\section{Introduction}
Cosmological inflation \cite{Guth:1980zm,Linde:1981mu,Albrecht:1982wi} provides a theoretically motivated paradigm for the origin of large scale structure, the origin of anisotropies in the Cosmic Microwave Background (CMB) and an explanation for a number of puzzles about the initial conditions for the 'Big Bang'.
In this paper, we present the background evolution of a lingering universe that leads to cosmological inflation -- emphasizing that this does not violate the Null Energy Condition (NEC).  In addition, we show that perturbations in the hydrodynamical fluids sourcing the background evolution grow during the lingering phase (which we will present in more detail in a second paper), before the transition to the inflationary phase.  Since Cosmic Microwave Background (CMB) measurements provide the amplitude of inhomogeneities at a given scale \cite{Akrami:2018vks}, any growth of perturbations in the lingering phase must be compensated by a corresponding decay in the period following the lingering.  It is important to note that since lingering demands the existence of positive curvature (i.e. the universe is closed), we will already have a lower bound on the duration of inflation.  The matching of inhomogeneities will provide a further restriction on the duration of lingering and inflation.

A lingering phase in cosmic evolution was first consider by Lemaitre \cite{10.1093/mnras/91.5.483} while investigating the dynamics of a closed universe with a cosmological constant.  Such a phase was reconsidered in more ``recent'' times to address the {\it age problem}, i.e. a discrepancy in determining the age of the universe when comparing observations of the Hubble to the age of the oldest globular clusters.  A similar type phase was suggested to provide a late-time mechanism for generating large scale structure (following matter domination) and could also lead to late-time cosmic acceleration \cite{Sahni:1991ks}. With improving observations and the successful predictions of inflation, the late lingering (loitering) universe fell out of favor.
However, as we discuss more below, such a phase is feasible for the very early universe and then leads to inflation.

\section{A Lingering Universe in Classical Gravity}\label{two}
We begin by considering the Friedmann, Lemaître, Robertson, and Walker (FLRW) metric for a homogeneous and isotropic universe
\be
ds^2=-dt^2 + a^2(t) \left( \frac{dr^2}{1-Kr^2} +r^2 d\Omega^2 \right),
\ee
allowing for spatial curvature $K$.
Introducing the Hubble parameter $H=\dot{a}/a$ and
allowing for a cosmological constant $\Lambda$
the Einstein equations can be written as
\bea
3H^2&=&\kappa^2 \rho -\frac{3 K}{a^2} +\Lambda, \label{hub}\\
\dot{H}&=&-\frac{\kappa^2}{2} \left(\rho +p\right)+\frac{K}{a^2}, \label{NECeom}
\eea
where $\kappa^2\equiv8\pi G$. Combining the above equations we get the Friedmann equation       
\be
\frac{\ddot{a}}{a}=-\frac{\kappa^2}{6}\left( \rho+3p\right)+\frac{\Lambda}{3}.
\ee
The energy density $\rho$ and pressure $p$ obey the continuity equation\footnote{If we think of the cosmological constant term as a fluid we would have $p_\Lambda=-\rho_\Lambda=-\Lambda/\kappa^2$ so that $\Lambda$ has units of mass squared. 
Similarly, we could think of spatial curvature (mathematically) as an energy density with pressure $p_K=-\rho_K/3=K/(\kappa a)^2$ and $\Omega_K=-K/(aH)^2$ is the critical density in spatial curvature. }
\be
\dot{\rho}+3H(\rho+p)=0.
\ee

The above equations were considered in various limits in the early days of cosmology to address observations of extragalactic `nebulae' (see Table I).
A particularly interesting account was given by Lemaître \cite{10.1093/mnras/91.5.483} to resolve an issue between models of de Sitter and Einstein.  The upshot of that work was that the universe must expand in the presence of matter and energy and these static solutions are unstable. 
In modern terminology we can recast \eqref{hub} as 
\be
\frac{H(t)^2}{H_0^2}=\Omega_{\Lambda,0} +\Omega_{r,0} \left(\frac{a_0}{a}\right)^{4} +\Omega_{m,0} \left(\frac{a_0}{a}\right)^{3}+\Omega_{K,0} \left(\frac{a_0}{a}\right)^{2}
\ee
where $H_0$ is a constant value for the Hubble parameter (typically taken as its value today) when $a = a_0$, and $\Omega_{i,0}$'s are the corresponding critical densities in the cosmological constant, radiation, matter, and curvature respectively.
These values have been determined with great precision \cite{Planck:2018vyg} demonstrating -- among other things --
that a constant value of the Hubble parameter is not consistent with General Relativity.

\begin{table}[t] 
\caption{Solutions for FLRW Cosmologies} 
\centering      
\begin{tabular}{|c |c |c |c|c|c|}  
\hline\hline
 a(t) & $H(t)$ & K & $\Lambda$ & Ricci Scalar & Description \\
\hline\hline                
$1$&0&\;\;\;\;0\;\;\;\;&\;\;0\;\;&0 & Minkowski $\left( {\cal{M}}^4 \right)$ \\ [0.4ex] 
\;\;$\exp(H_0 t)$ &\;\; $\sqrt{\Lambda / 3}$ \;\;&0 & $\Lambda$ &\;\; $4 \Lambda=12H_0^2$ \;\;& De Sitter (dS)   \\  [0.4ex] 
$a_0 $&0&$+$1&0& ${(6 K) }/{a_0^2}$  &    Einstein Static Universe \\ [0.4ex] 
$a_0$&$ 0$ &+1&$\Lambda$&${(6 K) }/{a_0^2}$  &\;\; \LM \, (a.k.a. lingering)\\  [0.4ex] 
\hline     
\end{tabular} 
\label{table:nonlin}  
\end{table} 

In the remainder of this paper we revisit some of these solutions. Firstly, it is important to note that some of these `states' for the universe can be realized as asymptotic fixed points of the dynamic flow from one state of the universe to another as time evolves -- we make this more precise below.  
As an example, if inflation is past-eternal it is typically assumed that this would lead asymptotically to dS space-time in the distant past. 
This would again imply that the universe began in a singularity (as in the standard Big Bang theory) as it is expected dS is geodesically incomplete in its infinite past implying the existence of a true singularity \cite{Borde:2001nh} -- for a more recent analysis we refer to \cite{Kinney:2023urn}. We also note that bouncing universes, such as the Cyclic / Ekpyrotic universes have also been shown to suffer from these issues \cite{Kinney:2021imp}. 

Moreover, many of the examples in Table I correspond to models with non-trivial spatial curvature. Spatial curvature is typically neglected in model building because the observationally determined value today from CMB, lensing, and BAO measurements is $\Omega_K = 0.0007 \pm 0.0019$ \cite{Planck:2018vyg}.
However, it is also interesting that in the 2018 Planck results  \cite{Planck:2018vyg} it was suggested that a non-zero spatial curvature might be favored by the data\footnote{There are many degeneracies in the data and we are in no way stating there is strong evidence for this. Further data will help clarify the situation.}.

\subsection{Dynamics in a lingering Universe \label{dynloit}}
In what follows it will be useful to work in conformal time $d\eta = dt / a(t)$ and we will consider matter and energy with equation of state
$p_i = w_i \rho_i$, where with $p_i$ is the pressure and $\rho_i$ is the energy density of the $i$-th fluid.
We will consider two types of fluids, one that is Standard Model-like $w_s \geq 0$, and another that scales slower than curvature with $w_e\leq-1/3$.
The equations for the background are
\begin{align}
\mathcal{H}^2 &= \frac{\kappa^2}{3} a^2 \rho_{tot} - K, \label{conFried1}\\
\mathcal{H}'+\frac{\mathcal{H}^2}{2} &= -\frac{1}{2}(\kappa^2 a^2 p_{tot} + K), \label{conFried2}
\end{align}
where $\mathcal{H}\equiv a'/a$ is the conformal Hubble parameter, 
 $\rho_{tot}=\sum_i \rho_i$ is the total energy density of the multiple sources and $p_{tot}$ is the corresponding total pressure.  We will find it more convenient to rescale our space-time co-ordinates by the spatial curvature. We can do this without changing the form of \eqref{conFried1} and \eqref{conFried2} since the FLRW metric is invariant under the scaling
\begin{align*}
\eta\rightarrow L\eta, &\quad \Vec{x} \rightarrow L\Vec{x} \\
K &\rightarrow L^{-2} K \\
a(\eta) &\rightarrow L^{-1} a(\eta).
\end{align*}
This means we can choose $L = \sqrt{K}$ such that we work with variables $\tilde{\eta} = \sqrt{K} \eta$, etc\footnote{Note that this means our scale factor now carries units of length, while our co-ordinates are unitless. Typically, one then uses the notation $R(\eta)$ for the scale factor, but we use $a$ for the sake of vexation.}.  We drop the tildes in what follows unless more context is necessary.  Solutions follow from solving \eqref{conFried1} and \eqref{conFried2} with $K=1$, along with the continuity equation for each fluid
\be
\label{BCons}
\rho_i'+3\mathcal{H}(1+w_i)\rho_i = 0.
\ee
Assuming a constant equation of state $w_i$ for each fluid the energy density scales as $\rho_i \sim a^{-3(1+w_i)}$. Using \eqref{conFried1} in \eqref{conFried2} we have
\be
\label{conFried2v2}
\mathcal{H}' = -\frac{\kappa^2 a^2}{6}\left(\rho_{tot} + 3 p_{tot}\right).
\ee
Another useful combination of \eqref{conFried1} and \eqref{conFried2} results in a second-order equation for the scale factor
\be
\label{Scale2}
a'' + a = -\frac{\kappa^2 a^3}{6}(\rho_{tot} - 3 p_{tot}).
\ee

\noindent {\bf The Lingering Fixed Point \label{loitloit}}
\begin{figure}
    \centering
         \includegraphics[width=0.6\textwidth]{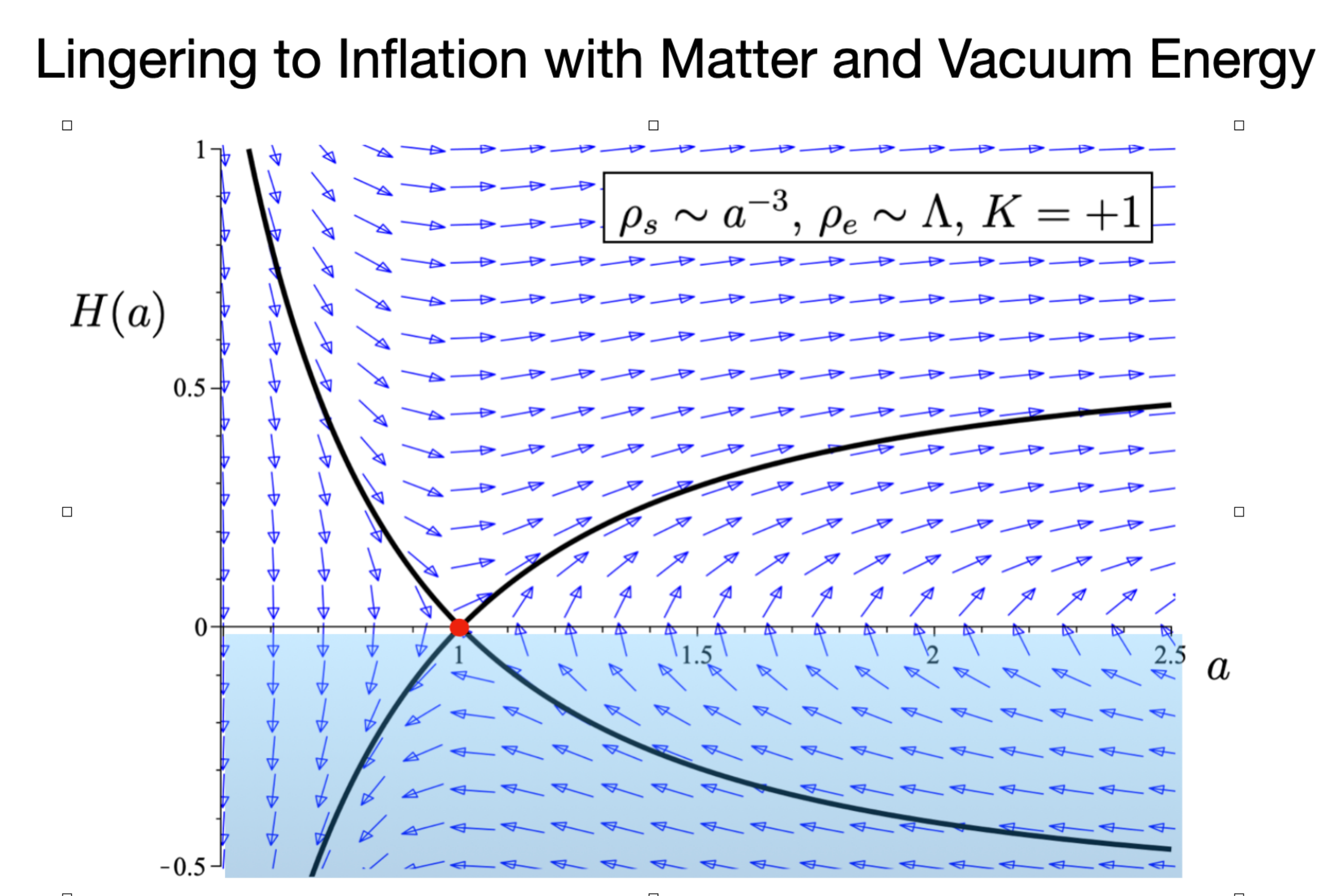}
    \caption{{ {\it Lingering Fixed Point.} The plot shows the flow lines in the Hubble parameter - scale factor phase space for a positively curved universe with standard matter and a fluid that scales like the cosmological constant. The red dot is an unstable (hyperbolic) fixed point in the evolution, it corresponds to initial conditions for exactly reaching the lingering phase. The black curves in both figures correspond to the enforcement of the Hubble constraint (energy conservation). } \label{fig:flow}}
\end{figure}

\begin{figure}[t]
         \centering
         \includegraphics[width=0.6\textwidth]{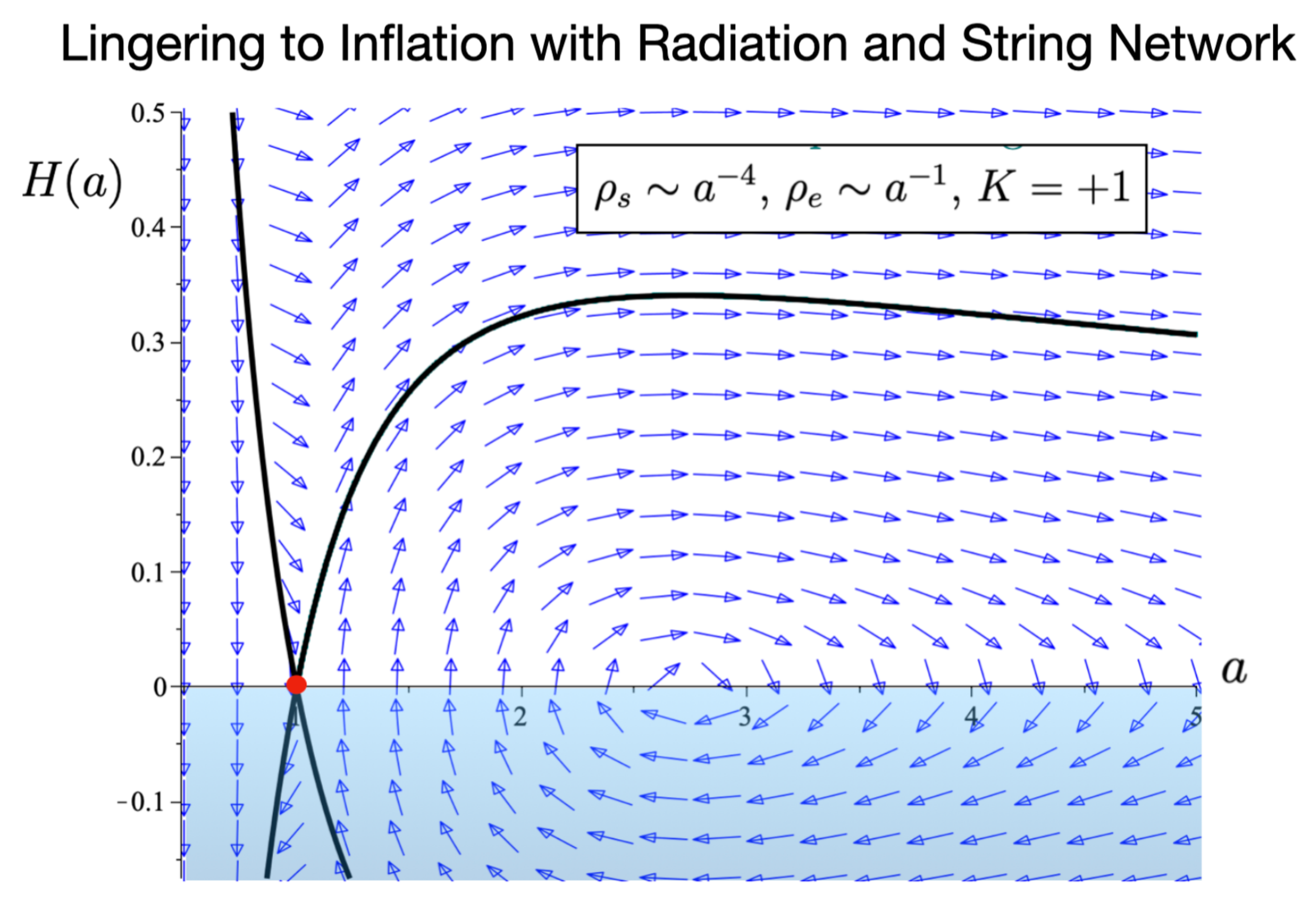}
    \caption{{ {\it Exact lingering.}
    The plot shows the flow lines in the Hubble parameter - scale factor phase space for a positively curved universe with radiation and a fluid that scales like a string network. The red dot is an unstable (hyperbolic) fixed point in the evolution, it corresponds to exact lingering. The black curve corresponds to the enforcement of the Hubble constraint (energy conservation). } \label{fig:flow_2}}
\end{figure}

A strictly lingering phase corresponds to a constant scale factor and vanishing Hubble parameter. While in conformal time we have $a'=a''=0$, the conformal Hubble radius $\mathcal{H}^{-1}$ diverges. Using \eqref{conFried2v2} and \eqref{Scale2}
a lingering phase implies
\begin{align}
\frac{\overline{\rho}_s}{a^m} + \frac{\overline{\rho}_e}{a^n} - \frac{3}{\kappa^2 a^2} = 0, \label{adot0} \\
(m - 2) \frac{\overline{\rho}_s}{a^m} + (n - 2) \frac{\overline{\rho}_e}{a^n} = 0, \label{addot0}
\end{align}
where $\overline{\rho}_s$ corresponds to standard matter and radiation, and $\overline{\rho}_e$ corresponds to an additional sector with $-1 \leq w_e \leq -1/3$, {\it which does not violate the NEC}. We use a bar to represent these quantities in the {\it strictly lingering phase}. 
For later convenience we have defined
$n\equiv3 + 3 w_e $ and $m\equiv 3 + 3 w_s $, which are related to the scaling $\rho_e \sim 1/a^n$ and $\rho_s \sim 1/a^m$ in the case of constant equation of state.  The constraints then imply $-2 \leq n \leq 2$ and $m \geq 3$. We emphasize that the conditions \eqref{adot0} and \eqref{addot0} are true for a strictly lingering universe where the first and second time derivatives of the scale factor vanish \textit{at all times}. 

As mentioned above, in \cite{Sahni:1991ks} the authors considered a period of cosmic lingering. They were seeking lingering solutions at low red-shift (late in matter domination) to address discrepancies in measured values of $H_0$ from CMB observations and those from clustering\footnote{It may be interesting to revisit this idea to address the current Hubble tension.}.  They also considered if their model could provide a possible mechanism for dark energy.  To do this they decided upon initial conditions based on observations
that would give appropriate amounts of spatial curvature, {\it exotic} matter, and clustering matter. Given these considerations they determined at which specific red-shift the universe must linger. 

Our approach here is very different -- we want to impose lingering as an initial condition for the beginning of inflation. Thus, rather than finding specific values of $\overline{\rho}_s$ and the scale factor such that the universe stalls, we determine the values of $\overline{\rho}_s$ and $\overline{\rho}_e$ that are allowed by a given scale factor and curvature. We find that for
\be
\label{LoitEnDens}
\overline{\rho}_s = \overline{\rho}_s^\ast \equiv \frac{2-n}{m-n} \frac{3}{\kappa^2} a_\ast^{m-2} \quad \mbox{and}, \quad \overline{\rho}_e = \overline{\rho}_e^\ast \equiv \frac{m-2}{m-n} \frac{3}{\kappa^2} a_\ast^{n-2},
\ee
the universe will linger at scale factor $a_\ast$. Notice that our assumptions on the equations of state imply that both energy densities remain positive $\bar{\rho}_i \geq 0$. 

We should immediately note that the $n=2$ case is special.  In that case, the universe can linger at \textit{any} scale factor as long as $\overline{\rho}_s = 0$ and $\overline{\rho}_e \sim constant$.
The resulting dynamics can be seen from the phase space plots in Figures \ref{fig:flow} and \ref{fig:flow_2}. The first figure illustrates the flow for a universe from lingering in the presence of matter and vacuum energy, asymptotically leading to inflation. Positive spatial curvature is essential for this solution to be consistent. Figure \ref{fig:flow_2} is an alternative situation where we consider radiation and a cosmic string network -- again leading to inflation. An important question we cannot answer is the duration of the lingering phase as we elaborate on below.


Our interest now lies in the amount of matter for which the lingering phase ends in a finite amount of time, i.e. $\overline{\rho}_s \neq \overline{\rho}_s^\ast$ and $\overline{\rho}_e \neq \overline{\rho}_e^\ast$.  We can characterize the length of the lingering phase by tracking ``nearby'' scale factor trajectories.  To this end, we suppose that $\overline{\rho}_e = \overline{\rho}_e^\ast(1 + \Delta_e)$, and $a(\eta) = a_\ast (1 + \Delta(\eta))$.  If the additional sector  $\overline{\rho}_e$ scales like curvature ($w_e=-1/3$ corresponding to $n=2$) we must remember that the lingering amount of clustering matter is zero.  In that case, we can't parameterize the deviation of the energy density as a fraction of what is required for lingering.  We must include two possible changes to the clustering matter: $\overline{\rho}_s = \overline{\rho}_s^\ast(1 + \Delta_s) + \Delta\rho_s$, where $\Delta_s$ will vanish in the $n=2$ case. Expanding  \eqref{conFried1} and \eqref{conFried2} to first order in the departure from lingering we have
\begin{align}
(m - 2) \Delta_e + (2 - n) \Delta_s &= 0, \label{Hub1O1} \\
\Delta'' + \frac{1}{2} \bigg( (m - 2) (n - 2) \Delta + \frac{(m - 2) (n - 3)}{m - n} \Delta_e - \frac{(m - 3) (n - 2)}{m - n} \Delta_s \bigg) &= 0. \label{Hub2O1}
\end{align}
The first of these equations ensures that $\kappa^2 a_\ast^2 \rho_{tot}/3 =  K = 1$. Assuming the new trajectory exactly matches the perfect lingering solution at some starting point $\eta_0$, the above equations are solved by
\be
\label{LoitScaleGen}
\Delta = \frac{\Delta_e}{2 - n} \bigg[ \cosh\bigg(\Delta\eta\sqrt{\frac{1}{2}(m-2)(2-n)} \bigg) - 1 \bigg],
\ee
where $\Delta\eta = \eta - \eta_0$. When $n=2$ the resulting equations are instead 
\begin{align}
\Delta_e + \frac{\kappa^2}{3 a_\ast^{m - 2}} \Delta\rho_s &= 0, \label{Hub1N2O1} \\
\Delta'' + \frac{1}{6}\bigg( \frac{\kappa^2}{a_\ast^{m - 2}} (m - 3) \Delta\rho_s - 3 \Delta_e \bigg) &= 0, \label{Hub2N2O1}
\end{align}
which is easily solved for
\be
\label{LoitScale2}
\Delta = \frac{m - 2}{4}\Delta_e \Delta\eta^2.
\ee
Note that one can pass from the $n\neq2$ solution to the $n=2$ solution via the formal limit $n\rightarrow 2$.

\begin{figure}[t]
    \centering
    \includegraphics[width=\textwidth]{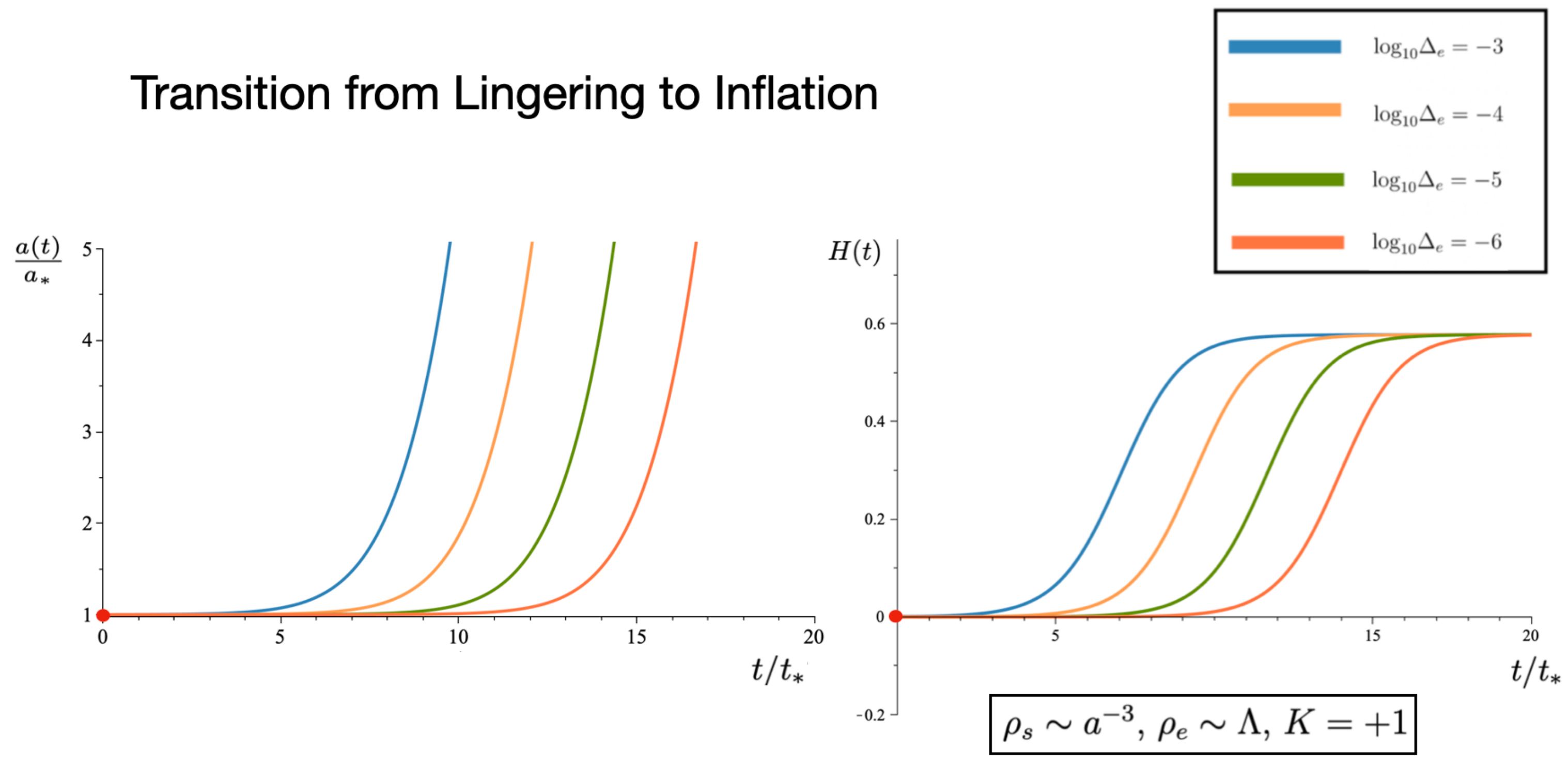} 
    \caption{
    { {\it Departure from lingering.} The left plot shows the cosmic-time evolution of the scale factor for different $\Delta_e$'s in a positively curved universe containing standard matter and a fluid that scales like the cosmological constant. A smaller $\Delta_e$ corresponds to a longer lingering period. The right plot shows the corresponding variation of the Hubble parameter, demonstrating the transition from lingering to dS.  
    The scale factor and time co-ordinate at the closest proximity to lingering are given by $a_\ast$ and $t_\ast$, respectively.  
    }
    \label{fig:my_label4}}
\end{figure}

Of course, this linear approximation breaks down when $\Delta$ approaches one; this signals the exit from the lingering phase.  Via inspection of the solutions above, one finds that with energy densities closer to their exact lingering values, the longer the duration of the quasi-lingering phase lasts. This is shown in fig. \ref{fig:my_label4}. We can estimate the end of lingering as
\be
\Delta \eta_{e} \simeq \sqrt{\frac{2}{(m - 2) (2 - n)}} \ln \bigg((2 - n) \frac{2}{\Delta_e} \bigg)
\ee
when $n < 2$. We assumed that $\Delta_e \ll 1$ to arrive at the above expression. When $n = 2$, 
\be
\Delta \eta_{e} = \frac{2}{\sqrt{(m - 2) \Delta_e}}. 
\ee
Both of these time durations obey $\Delta\eta_e\rightarrow\infty$ for $\Delta_e\rightarrow0$, which is the perfectly lingering limit.


To this point, we made no assumptions on $\Delta_e$, $\Delta_s$, or $\Delta\rho_s$. The forms of \eqref{Hub1O1} and \eqref{Hub1N2O1}, as well as the interesting ranges for $m$ and $n$, require the changes in the energy density of the fluids to be opposite-signed. The universe, desiring to stall at the same scale factor as in its perfect lingering form, must compensate any additional energy in a given fluid by a corresponding loss in the other. Simple inspection of the solutions to the deviated lingering equations reveals that when we have reduced the amount of energy in the additional sector, the universe tends to recollapse, as expected for a closed universe filled with standard matter.

The $n=2$ case introduces further complications that necessitate extra elucidation. Note that the perfect lingering scenario requires zero standard matter to achieve a steady-state universe. Thus, any additional matter sector inserted to break lingering requires the corresponding reduction of standard matter. In other words, the universe needs negative energy density in the standard matter sector to obtain an initially lingering solution followed by an exponentially expanding inflation-like solution.

We note that these considerations qualitatively and quantitatively change the nature of the post-lingering behavior of the scale factor. Below, we consider the universe to be dominated by the additional sector fluid after lingering, a fact which depends on the additional sector diluting slower, with cosmic expansion, than the standard matter. If, instead of continuing to expand, the universe shrinks, the reverse situation holds. With cosmic contraction, standard matter concentrates faster than the additional sectors. It would only be a short time before the standard matter dominated over the additional sector. Since our current universe appears to be expanding at an accelerated rate, we ignore the options of collapsing universe in what follows.

\subsection{Post-lingering and the Exit to Inflation}\label{three}

\begin{figure}
    \centering
    \begin{subfigure}[b]{\textwidth}
         \centering
         \includegraphics[width=0.6\textwidth]{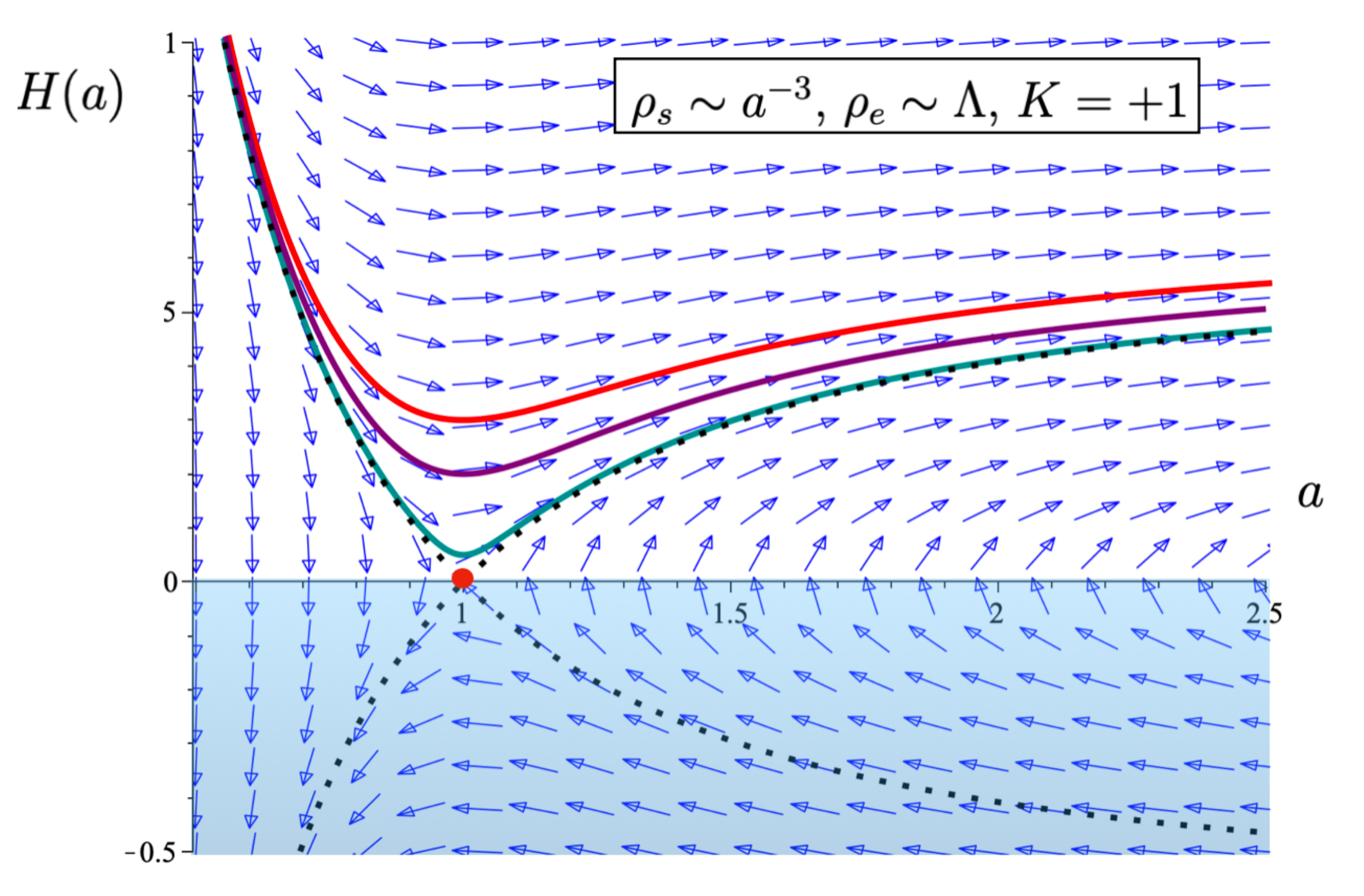}
    \end{subfigure}
    \caption{  {\it Exiting to Inflation.} The plot shows the flow lines in the Hubble parameter - scale factor phase space for a positively curved universe with standard matter and a fluid that scales like the cosmological constant. The dotted trajectory  corresponds to the enforcement of the Hubble constraint with $\Delta_{e} = 0$, it passes through the red dot which corresponds to the fixed point of lingering. The colored curves are trajectories with $\Delta_{e} \neq 0$, which correspond to an exit from the lingering phase to inflation. Further away a curve is from exact lingering, the larger the value of $\Delta_{e}$. \label{fig:flow2}}
\end{figure}

By their scaling, matter and radiation dilute with the cosmic expansion faster than the dS-like sector.  Neglecting the matter sector then provides us with the equations that determine the evolution of the scale factor in the post-lingering era.  In this limit, \eqref{Scale2} takes the form
\be
\label{PLDifEq1}
a'' + a = \frac{(m-2)(n-4)}{(n-m)} \frac{(1+\Delta_e)}{2 a_\ast^{2-n}} a^{3-n} \equiv C a^{3-n},
\ee
where we have inserted the ansatz for $\overline{\rho}_e$ discussed above.  The first Friedmann equation then takes the form
\be
\label{PLDifEq2}
\Hcon^2 = \frac{2(n-m)C}{(3-n)(m-2)(n-4)} a^{2-n} - 1.
\ee
The above equations can be solved for $a(\eta)$ for the relevant values of $n$. We denote by $\Delta \eta_{pl}$ the conformal time duration since the post-lingering transition. The general solution for \eqref{PLDifEq1} and \eqref{PLDifEq2} is
\be
\label{PostLoitScaleGen}
a_n = \Bigg(-\frac{m-2}{n-m} \frac{1+\Delta_e}{a_\ast^{2-n}}\Bigg)^{1/(n-2)} \sin(\frac{n-2}{2}\Delta\eta_{pl} + \eta_{1})^{2/(n-2)},
\ee
where $\eta_{1}$ is the remaining constant of integration set by initial value of the scale factor at the start of the post-lingering phase of evolution \ From this solution, it is clear that we need to treat the $n=2$ case separately. For $n=2$, using the exponentially expanding solution of \eqref{PLDifEq1} in \eqref{PLDifEq2}, we find
\be
\label{Scalen2}
a_2 = a_{pl} \exp\left(\Delta \eta_{pl} \sqrt{\Delta_e}\right),
\ee
where  $a_{pl}$ is the scale factor at the the lingering to post-lingering transition.


\section{Cosmology from a Hagedorn Phase  \label{stringcosmo}}
In this section, we present the motivation from string theory for an initial phase of cosmological lingering.
We then demonstrate how this can lead to a period of inflation without a NEC violating transition. 
In regards to the presentation thus far, within the context of string theory (quantum gravity) we address the question:
{\it Is spatial curvature an essential component for a NEC preserving, lingering universe\footnote{{\bf Spoiler Alert:} Our results respect Betteridge's law of headlines.}?} We also discuss how our approach relates and differs from previous considerations in the literature. 

\subsection{Model Independent String Cosmology}
Essentially all string theory compactifications give rise to the following terms in the effective action (written in the four-dimensional string frame) \cite{Polchinski:1998rq,Green:1987sp,Green:1987mn}:
\bea \label{stringaction}
S&=&\frac{1}{2 \kappa_4^2} \int  d^4x  \sqrt{-G} e^{-\phi_s} \left[ R + (\partial \phi_s)^2 -\frac{1}{12} H_3^2\right] + \frac{1}{2} \int  d^4x \; {\cal L}_m\left(\phi_s,G_{\mu \nu}, \ldots\right), \\ \nonumber
e^{-\phi_s}&=& \frac{\mbox{Vol} \, {\cal{M}}_6}{(2\pi \sqrt{\alpha^\prime})^6} \frac{1}{g_s^2}, \bigskip 2\kappa_4^2=2\pi \alpha^\prime \equiv \frac{\textcolor{blue}{1}}{M_s^2},
\eea
 where (for simplicity) we have assumed a string compactification from ten to four dimensions ($\mbox{Vol} \, {\cal{M}}_6$ is the volume of the extra dimensions in string units), $G_{\mu \nu}$ is the string frame metric, $R$ is the corresponding Ricci scalar, $g_s$ is the string coupling, and $M_s$ is the string scale. We have focused on the bosonic string spectrum assuming weak coupling and small space-time curvature. We have also neglected fermions, tachyons, critical dimensions, world-sheet curvature, Ramond - Ramond fluxes, and the possibility of supersymmetry as {\it including these effects would not alter our main conclusions}. However, we have included a term ${\cal L}_m$ which can account for arbitrary sources of the dilaton. 
Finally, $H_3$ is the three-form flux (an Einstein - Maxwell field) and $\phi_s$ a scalar field -- the dilaton.
 
 Given our analysis in the previous sections, it will be useful to recast this theory in the Einstein frame.
 We emphasize that it is in the Einstein frame that one should invoke the NEC if one is interested in the stability of the theory and the presence of singularities. There is a notion of these issues in the string frame (physics does NOT depend on the frame, however the singularity theorems were originally formulated in the Einstein frame \cite{Hawking:1973uf}, and are most easily interpreted in the context of General Relativity\footnote{One can consider the NEC in both frames but it is simpler to interpret in the Einstein frame \cite{Gasperini:1992em} where violations can be interpreted through energy and matter following geodesics in a clear way as originally stated by Hawking and Ellis \cite{Hawking:1973uf}.}.
 In the four-dimensional Einstein frame, defined by the conformal transformation (which does not affect experimental observations) 
 \bea \label{conformtrans}
 ds_{s}^2\rightarrow ds_E^2 = e^{-\phi_s} ds_{s}^2, \bigskip
 G_{\mu \nu} \rightarrow g_{\mu \nu}= e^{-\phi_s} G_{\mu \nu}, \bigskip 
 \phi_s\rightarrow\phi_{E}=\phi_s 
\eea
 \eqref{stringaction} becomes
\bea \label{eframeaction}
  S =\int d^4 x \, \sqrt{g}
   \left[  \frac{R[g_{\mu \nu}]}{2\kappa^2}  - {\frac{1}{2}} (\partial\phi_E)^2 - \frac{1}{2} 
    e^{2\phi_E} (\partial\chi)^2 \right] + \frac{1}{2} \int  d^4x \; {\cal \tilde{L}}_m\left(\phi_E,g_{\mu \nu}e^{\phi_E}, \ldots\right),
 \eea 
where $\kappa^2 = 1/M_p^2$ as in previous sections, and the fields $\phi_E$ and $\chi$ have been rescaled, the gravitational term has the usual Einstein-Hilbert form, and the dilaton source ${\cal \tilde{L}}_m$ now has dilaton dependent coupling. Except in Type I string theory, $H_3$ is the four-dimensional part of the field strength of the Neveu-Schwarz two-form, and $\chi$ is defined via the equation $d\chi = e^{-\phi_s} *H_3$.
The dilaton $\phi_s$ and the axion $\chi$ are common to nearly all string theories
and this simple model captures many aspects that string compactifications and the resulting cosmologies have in common. We might immediately suppose that there are FLRW solutions where $\chi + i e^{-\phi_s}$ approximately describes a semi-circular geodesic in the upper half plane (with a relevant field-space metric), and that by making $g_s$ small, we can avoid significant back-reaction of the scalars on the geometry.  
Again, the action becomes more involved in the presence of supersymmetry (and its breaking), additional fields, branes, extra dimensions, etc. -- but these can (and have) been accounted for in the literature to a large extent (see e.g. \cite{Silverstein:2004id,Kane:2015jia}). Here we will focus on the $3+1$ dimensional cosmologies as we did above, and in the simplest terms to reach our conclusions. 

The role of $\hbar$ in string theory is played by the Vacuum Expectation Value (VEV) of the dilaton, which is related to the string coupling $g_{s}^{2} \sim \exp \langle \phi_s \rangle$ and in the dimensionally reduced theory this would also include the extra dimensions as expressed in \eqref{stringaction}. The the role of gravitational corrections are controlled by the string scale $M_s$ -- which here we will take parametrically below the Planck scale ($1/\kappa = M_p = e^{\phi_s/2}M_s$). In particular, the Planck scale is a derived quantity which involves: the dynamics of the dilaton (its resulting VEV), the value of the string scale, and the size and dynamics of any extra dimensions (here assumed six-dimensional\footnote{Strictly speaking, string theories do not require 10 or 11 dimensions to be consistent, nor do they require supersymmetry \cite{Polchinski:1998rq,Green:1987sp,Green:1987mn}. This may be an important clue on how to realize realistic models of cosmology from string theory -- such as quasi- but not exact dS space-time \cite{Hellerman:2001yi}.} ${\cal M}_6$). An important consequence of this is that when one considers decoupling limits one must take into account the other scales (and dynamics) involved. 
Looking into these consequences has led to interesting conjectures -- such as the Weak Gravity conjecture and notion of the Swampland of quantum field theories \cite{Ooguri:2006in,Vafa:2005ui,Ooguri:2018wrx,Garg:2018reu}.

Moreover, string theory comes with a number of dualities \cite{Polchinski:1998rq,Green:1987sp,Green:1987mn} resulting from the worldsheet theory and manifested in the low-energy action \eqref{stringaction}.  These symmetries can be used to relate the weak coupling limit {\it of one realization of the theory} with the strong coupling of another. When applied to the geometry they also imply a theoretical correspondence between the large volume limit {\it of one theory} with the small volume limit {\it of another theory}. Finally, when applied to cosmology this would imply a `scale-factor' duality \cite{Brandenberger:1988aj,Veneziano:1991ek,Tseytlin:1991xk}. Given the latter, string theory was used to motivate alternatives to standard inflation
resulting from a bouncing universe -- examples include String Gas Cosmology (SGC) \cite{Brandenberger:1988aj,Battefeld:2005av} and the Pre-Big-Bang (PBB) \cite{Gasperini:1992em}. 

{\bf What we are not considering in this paper.} {\it In SGC and PBB models, the goal is to produce {\bf an alternative to standard inflation} to resolve the classic issues of the background (the Horizon, flatness, entropy problems, etc.), while also (causally) generating cosmological perturbations to provide seed perturbations for the CMB and Large-scale Structure\footnote{One possible exception to this appears in \cite{Kamali:2020drm}. However, our approach differs in that those authors again considered a {\bf bouncing cosmology} in the context of SGC that then led to inflation.}. One challenge for these approaches is that they invoke a `bounce' before or near the cosmological singularity, and in perturbation theory (required to trust the effective action \eqref{stringaction}) this implies a violation of the NEC. Also, as mentioned above such approachs can also suffer from singularity issues as established in \cite{Borde:2001nh,Kinney:2023urn,Kinney:2021imp}.  We will return to these points, but first we further discuss the motivation for a new phase of early universe cosmology -- a lingering phase. }

\subsection{The Hagedorn Phase}
In addition to \eqref{stringaction}, another robust prediction of string theories is their thermodynamic properties at high energy.
An important first step for understanding the high temperature behavior resulted from considerations of Hagedorn while 
exploring quark deconfinement and its dependence on temperature in \cite{Hagedorn:1965st}.
In particular, this study led Hagedorn to the idea of a cosmic and phenomenological limiting temperature. 
Such thermodynamical behavior signals a departure from standard QFT, the partition function will have an exponentially diverging degeneracy of states (hence the mass spectrum) at high energy and temperatures.
When interpreted within string theory, this implies a maximum temperature $T_H$ for a fluid of strings --  the so-called Hagedorn temperature \cite{Atick:1988si}.

Cosmologically, the epoch in which the limiting temperature is achieved, $T \approx T_H$, 
implies a phase where the temperature remains roughly constant -- a result from the fact that the energy of a string does not increase \cite{Atick:1988si}.
Naively, for FLRW cosmology this would imply that a gas of strings would evolve like pressure-less matter -- $\rho_H \sim E / a^3$ with $E$ constant.
However, an important consideration is the string dilaton, which would also be present within string theory \cite{Tseytlin:1991xk}.
The resulting dynamics implies a departure from standard FLRW intuition as manifested in the String frame \eqref{stringaction}.
Moreover, in addition to the Hagedorn phase and the dilaton, string theory also generically leads to extra dimensions and `winding strings'.
In a simple realization, we can consider extra dimensions in the form of a six-dimensional torus (a simple example of a Calabi-Yau manifold) with strings wrapping the cycles of these dimensions\footnote{We note that this may seem rather exotic to the unfamiliar, however the extra dimensions can also be used to understand important problems in particle phenomenology and cosmic inflation -- where it has been crucial in providing a systematic way to calculate corrections to the low-energy effective field theory.} in addition to unwrapped strings (particles -- the low energy states being the graviton, Einstein-Maxwell fields (model independent axions), and the dilaton). A cosmological consequence of the dilaton and the wrapped states is they don't lead to inflation (as in standard FLRW) but instead preserve the Hagedorn phase dynamics. 

To study the dynamics we start by considering a homogeneous but anisotropic metric
\begin{equation}
    ds^2 = n(t)^2 dt^2 + \sum_{i=1}^{N} e^{\lambda^{i}_s(t)}dx_{i}^{2}
\end{equation}
where $n$ is the lapse function which we gauge fix to 1 in what follows, $a^{(i)}_{s} = e^{\lambda^{i}_s(t)}$ are string frame scale factors for the $N$ spatial directions. Specializing to isotropic cosmologies ($\lambda^{i}_s(t) = \lambda_s(t) \ \forall \ i$) with $N=3$, the equations of motion resulting from \eqref{stringaction} are
\bea \label{eoms_string}
\dot{\phi}_s&=&3 {H}_s \pm \sqrt{3 {H}_s^2  + e^{\phi_s} \rho_s}, \nonumber \\
\dot{H}_s&=&\pm{H}_s \sqrt{3 {H}_s^2+e^{\phi_s} \rho_s} + \frac{1}{2} e^{\phi_s}  \left( p_s +\Delta_\phi {\cal L}_m \right), \nonumber \\
\dot{\rho}_s&+&3 {H}_s \left( \rho_s + p_s \right)=-\dot{\phi}_s \Delta_\phi {\cal L}_m,
\eea
where $\rho_s$ and $p_s$ are the string frame energy density and pressure, $H_s$ is the string scale Hubble parameter, and $\Delta_\phi {\cal L}_m$ is the variational derivative of ${\cal L}_m$ with respect to the dilaton, which we set to zero for now. We work in units $2\kappa_4^2=1$.
We note that these equations become the same as the Einstein frame equations in Section \ref{two} if
when we take the (unshifted) dilaton to be constant and include spatial curvature through the energy density. 
The energy and pressure of sources are defined in the string frame and given by\footnote{See \cite{Kaloper:2007pw} for more detail.} the comoving free energy as 
\be \label{EandP}
E_s = - (F + \beta \partial_\beta F)  \bigskip 
P = \partial_{\lambda_s} F = -\partial_{\lambda_s }E \left\vert_{\beta = {\mbox{\tiny const}}} \right.,
\ee
where $\beta$ is the inverse temperature and $\lambda_s=\ln a_s$. The sources obey the conservation equation
\be
\dot{E}_s+3 H_s  P=0.
\label{conseq}
\ee
We will focus on the
sources of the form $E_s= E_{0} \, e^{-3 \gamma \lambda_s}$
with equation of state parameter
$\gamma=P_s/E_s=p_s / \rho_s$, $\rho_s$ and $p_s$ are the energy and pressure density derived from the comoving quantities $E_s$ and $P_s$ respectively.
Given the considerations above, we will be interested in two cosmological phases.
One phase corresponds to the Hagedorn phase of strings where their energy does not increase and the temperature remains nearly constant.
This implies an equation of state for the Hagedorn phase -- $\gamma = 0$.
Whereas, below the Hagedorn phase (but still at relatively high temperatures $T \lesssim T_H$) strings have two types of scaling behavior\footnote{One may ask about the importance of string oscillations, but it has been shown that these can be neglected relative to the sources we consider \cite{Kripfganz:1987rh}.} corresponding to 
radiation ($\gamma = 1/3$) and winding strings ($\gamma=-1/3$).
In both epochs we will see the dilaton plays a crucial role in the dynamics. 

\subsubsection{Cosmological Solutions}
Solutions of the system \eqref{eoms_string} can be found 
for constant equation of state by introducing a time $x$ such that 
\be
dx=\left( \frac{E}{E_0} \right) dt, \label{solntime}
\ee
where $E_0$ is a constant reference energy. 
Using \eqref{solntime} the equations \eqref{eoms_string} are equivalent to 
\bea
\vps_s - 3 \lps_s&=& E_0 e^{\varphi_s+\gamma 3 \lambda_s}, \label{xeqns} \\
\vpp_s-3\gamma  \vp_s \lp_s -3 \lps_s &=& \half   E_0 e^{\varphi_s+3\gamma  \lambda_s}, \nonumber \\
\lpp_s-3\gamma  \lps_s - \vp_s \lp_s &=& \half \gamma E_0 e^{\varphi_s+3\gamma \lambda_s}, \nonumber
\eea
where primes denote derivative with respect to $x$, $\varphi_s = \phi_s - 3\lambda_s$ is the shifted dilaton, and energy scales as $E=E_0 \exp \left(  -3\gamma \lambda_s \right)$. 

Looking for isotropic solutions with constant equation of state $p=\gamma \rho$ one finds \cite{Gasperini:1993hu},
\bea
\lambda_s&=& \lambda_{s0}+ \frac{\gamma}{\alpha} \ln \left[ (x-x_-)(x-x_+) \right]+ \frac{1}{\alpha \sqrt{3}} \ln \left( \frac{x-x_+}{x-x_-} \right), \label{1nlos} \\
\varphi_s&=&\varphi_{s0}-\frac{1}{\alpha} \ln \left[ (x-x_-)(x-x_+) \right]-\frac{\gamma \sqrt{3}}{\alpha} \ln \left( \frac{x-x_+}{x-x_-} \right), \\
\phi_s&=&\phi_{s0}- \ln \left[  (x-x_-)(x-x_+) \right] - \frac{(\gamma-1)\sqrt{3}}{\alpha} \ln \left(  \frac{x-x_+}{x-x_-} \right), \label{3nlos} 
\eea
with singularities at
\be
x_{\pm}=\frac{1}{\alpha} \left[ 3\gamma x_1 - x_0 \pm \sqrt{ \left( 3\gamma x_1-x_0 \right)^2 + \alpha \left( 3x_1^2-x_0^2 \right)} \, \right],
\ee
where $\alpha=1-3\gamma^2$ and $\lambda_{s0}$, $\varphi_{s0}$, $\phi_{s0}$, $x_0$, and $x_1$ are integration constants. $x_-<x<x_+$ is a classically forbidden region. 

\subsubsection{Hagedorn Phase at $T \simeq T_H$}
For the Hagedorn phase with $T \simeq T_H$, the energy density is dominated by a fluid with constant energy $E=E_0$ and vanishing pressure $P=0$, so that $\gamma=0$ $(\alpha=1)$, in which case equations \ref{1nlos}-\ref{3nlos} become
\bea
\lambda_s&=& \lambda_{s0}+  \frac{1}{\sqrt{3}}  \ln \left( 1-\frac{x_*}{x} \right), \\
\varphi_s&=&\varphi_{s0} -  \ln \left[ x( x-x_*) \right],  \\
\phi_s&=&\phi_{s0}+  {\sqrt{3}}  \ln \left( 1-\frac{x_*}{x} \right)- \ln \left[ x( x-x_*) \right],
\eea
where we have set the integration constants so that $x_-=0$ and $x_+=x_*$. Using (\ref{solntime}), the solution in co-ordinate time is obtained by replacing $x$ with $t$ -- 
\bea
\lambda_s&=&\lambda_{s0}+\frac{1}{\sqrt{3}} \ln\left( 1-\frac{t_*}{t} \right), \label{sol1} \\
\varphi_s&=&\varphi_{s0}- \ln\left[  t \left(t-t_* \right) \right], \\
\phi_s&=&\phi_{s0}-2\ln t+\left( \sqrt{3}-1 \right) \ln \left(  1-\frac{t_*}{t} \right),
\label{sol2} 
\eea
where $t_* = x_*$ (again, the moment of the singularity) and the string coupling is
\be
g_s^2=e^{\langle \phi_s \rangle}=e^{\phi_{s0}} t^{-2} \left( 1-\frac{t_*}{t} \right)^{\sqrt{3}-1}.
\ee

From \eqref{eoms_string}, we see there are two branches of solutions --
\bea
\dot{\varphi}_s&=&\pm \sqrt{3 H_s^2 + e^{\varphi_s} E_0}, \label{sln1}\\
\dot{H}_s&=&\pm H_s  \sqrt{3 H_s^2+e^{\varphi_s} E_0}, \label{sln2} 
\eea
corresponding to the sign of $\dot{\varphi}_s$. Differentiating equations \ref{sol1} - \ref{sol2} gives
\bea
\dot{\varphi}_s&=&  - \frac{(2t-t_*)}{t(t-t_*)},\\
H_s&=& \frac{t_*}{\sqrt{3}} \left( \frac{1}{t(t-t_*)} \right).
\eea
As emphasized in \cite{Kaloper:2007pw}, these solutions represent different superselection sectors of the theory and are topologically distinct solutions categorized by the sign of $\dot{\varphi}_s$.
The solution branches are summarized in Table \ref{SolutionTable} where we present the 
four physically distinct solutions.  
\\

\begin{table}[t]
\begin{center}\begin{tabular}{|c|c|c|c|c|}
\hline   Region& Branch   &Expansion  & \;\;Shifted Dilaton\;\;  &  Dilaton  \\
\hline
I&$(+)$ &$H_s>0$ & $\vpd_s>0$ &  $\pd_s>0$  \\
II& $(+)$ &  $H_s<0$ & $\vpd_s>0$ &  $\pd_s>0 \; {\text or} \; \pd_s<0$  \\
 III&$(-)$ & $H_s>0$ & $\vpd_s<0$ & $\pd_s>0 \; {\text or} \; \pd_s<0$\\
 IV&$(-)$ & $H_s<0$ & $\vpd_s<0$ & $\pd_s<0$   \\
\hline \end{tabular} 
\caption{The four regions of the Hagedorn solution (as discussed in the text and Figure \ref{SGCfig}).  \label{SolutionTable} }
\end{center}
\end{table}

\noindent {\bf Phase Space Analysis of the Hagedorn Phase.}
Here we want to understand the phase space of solutions for the Hagedorn Phase.
To do this it is useful to introduce a co-ordinate transformation 
\be
d\tau=\sqrt{E_0}e^{(\varphi_s-3\gamma  \lambda_s)/2} dt, \label{tautime}
\ee
(Note: in this section primes represent the derivative with respect to $\tau$).
Then the equations are 
\bea
\vps_s-3 \lps_s =1,\label{tautimeeoms}   \\
\vpp_s+\half \vp_s(-3\gamma \lp_s - \vp_s)=-\half , \nonumber  \\
\lpp_s+\half \lp_s(-3\gamma  \lp_s - \vp_s)=\frac{\gamma }{2 }. \nonumber
\eea

Introducing $l(\tau)=\lp_s$ and $f(\tau)=\vp_s$ the above equations can be taken as a first order system with fixed points
\be \label{tautimefix}
\left( l,f \right)=\left(  \pm \sqrt{\frac{\gamma^2 }{(1-3\gamma^2 )}}, \pm sgn(-\gamma) \sqrt{\frac{1}{(1-3\gamma^2 )}}   \right),
\ee
where $sgn(\gamma)= \pm 1$ is the sign of $\gamma$. Near the critical temperature $\gamma = 0$ as discussed earlier.

\begin{figure}[t]
    \centering
         \includegraphics[width=0.65\textwidth]{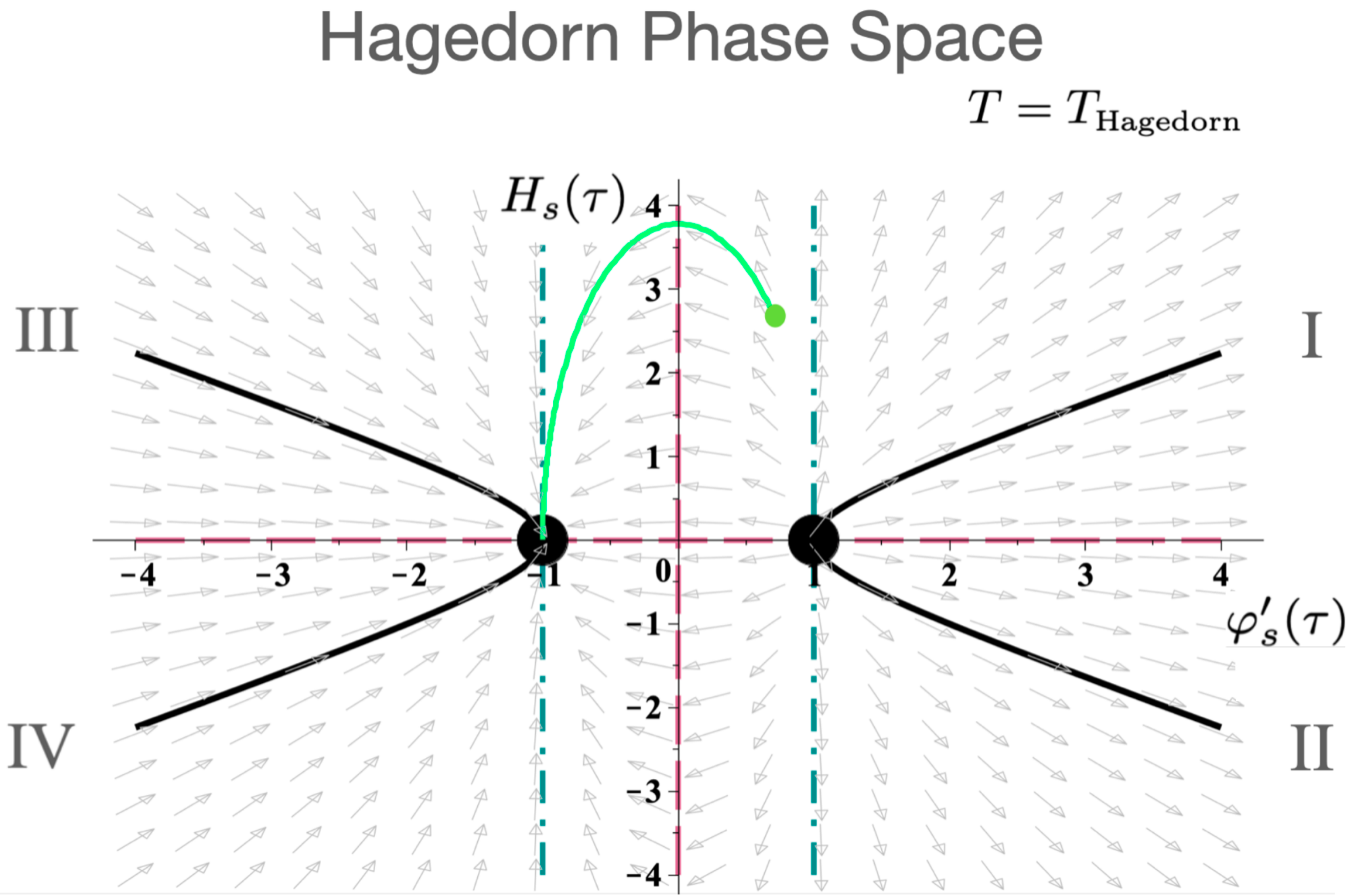}
    \caption{{\smaller 
        The phase space ($\varphi_s^\prime,H_s$) for Hagedorn cosmologies at the critical temperature. The black curves enforce the Hamiltonian constraint. The green trajectory is an example of an unphysical observer.  The blue (dot-dash) trajectories are lines of constant $\varphi_s^\prime$, whereas pink (dashed) trajectories correspond to constant $H_s(\tau)$.  Their intersection occurs at the causal fixed points of the phase space flow $\left(\pm 1,0 \right)$. 
    \label{SGCfig} }}
\end{figure}

 The phase space for Hagedorn cosmologies near the critical temperature $T_H$ is given in Figure \ref{SGCfig}.
 There we plot the string frame Hubble parameter in the conformal time $\tau$ versus the derivative of the shifted dilaton. 
   All physically relevant trajectories are restricted by the Hamiltonian constraint (conservation of energy) to begin and end on the black (co-dimension one) hyperbolae. This is an important observation in that a trajectory that start in region I or II can not energetically go to region III or IV.
  We performed the conformal transformation to make this manifest (we also note this transformation is not singular). 
  Thus, there are two distinct branches again separated by the sign of the shifted-dilaton. 
  The green trajectory provides an example of an (unphysical) observer that respects the phase space flow but does not satisfy conservation of energy.  The blue (dot-dash) trajectories are lines of constant (in conformal time) $\varphi_s^\prime$, whereas pink (dashed) trajectories correspond to constant (in conformal time) $H_s(\tau)$.  Their intersection occurs at the causal (hyperbolic) fixed points of the phase space flow $\left(\pm 1,0 \right)$. In summary,  
our results agree with those of \cite{Kaloper:2007pw}.

In appendix \ref{app1} we discuss how the Hagedorn phase in the string frame relates to that in the Einstein frame.

\subsection{Post-Hagedorn Phase $T \lesssim T_H$}
We now turn to the cosmology when the temperature has decreased below the Hagedorn temperature.
Such behavior is expected due to the fact that the lingering solution of the Hagedorn phase is a hyperbolic (unstable) fixed point for $\varphi^\prime_s >0$ as can be seen from Figure \ref{SGCfig}. We will see in the next section that these solutions are the ones of interest for lingering cosmologies that respect the NEC. 
As mentioned above the {\it stringy} nature of particles will manifest itself as both radiation and string winding modes below the Hagedorn temperature -- and the dilaton will play a crucial role until stabilized.
From the discussion above, we can use the solutions from 
\eqref{1nlos}-\eqref{3nlos} where we have $\gamma= +1/3  \; (-1/3)$ for radiation (winding modes). 
The solutions are
\bea
\lambda_s&=& \lambda_{s0} \pm \frac{1}{2} \ln \left[ x(x-x_*) \right] + \frac{\sqrt{3}}{3} \ln \left( 1-\frac{x_*}{x} \right), \label{radsoln1}\\
\varphi_s&=&\varphi_{s0} -\frac{3}{2} \ln \left[ x(x-x_*) \right] - \frac{\sqrt{3}}{2}  \ln \left( 1-\frac{x_*}{x} \right),  \\
\phi_s&=&\phi_{s0}+\sqrt{3}  \ln \left( 1-\frac{x_*}{x} \right), \label{radsoln3}
\eea
where again we set $x_-=0$ and $x_+=x_*$. The upper sign in \eqref{radsoln1} corresponds to radiation (momentum modes) and the lower sign to winding modes. 
When considering the behavior far from the singularities (i.e. $|x| \rightarrow \pm \infty$), the energy for the momentum modes is given by $E \sim e^{-\lambda_s}$, and noting the relation to the original co-ordinate time,  we find that
$|x| \sim |t|^{1/2}$ and the solutions {\it in this limit} approach
\bea
\lambda_s&\rightarrow& \lambda_{s0}+  \frac{1}{ 2} \ln ( t), \\
\varphi_s&\rightarrow&\varphi_{s0} -\frac{3}{2} \ln (t),  \\
\phi_s&\rightarrow&\phi_{s0},
\eea
which is the standard FLRW radiation dominated universe.
In general,  the radiation phase leads to four physically distinct solutions which are expressed in Table \ref{RadiationTable}. Likewise, we can find the solutions for the winding strings using T-duality, and oscillations of the strings can be neglected cosmologically. 
For more details we refer the reader to \cite{Kaloper:2007pw}.
Here we want to emphasize the analysis of the phase space.

\begin{figure}[t]
    \centering
         \includegraphics[width=\textwidth]{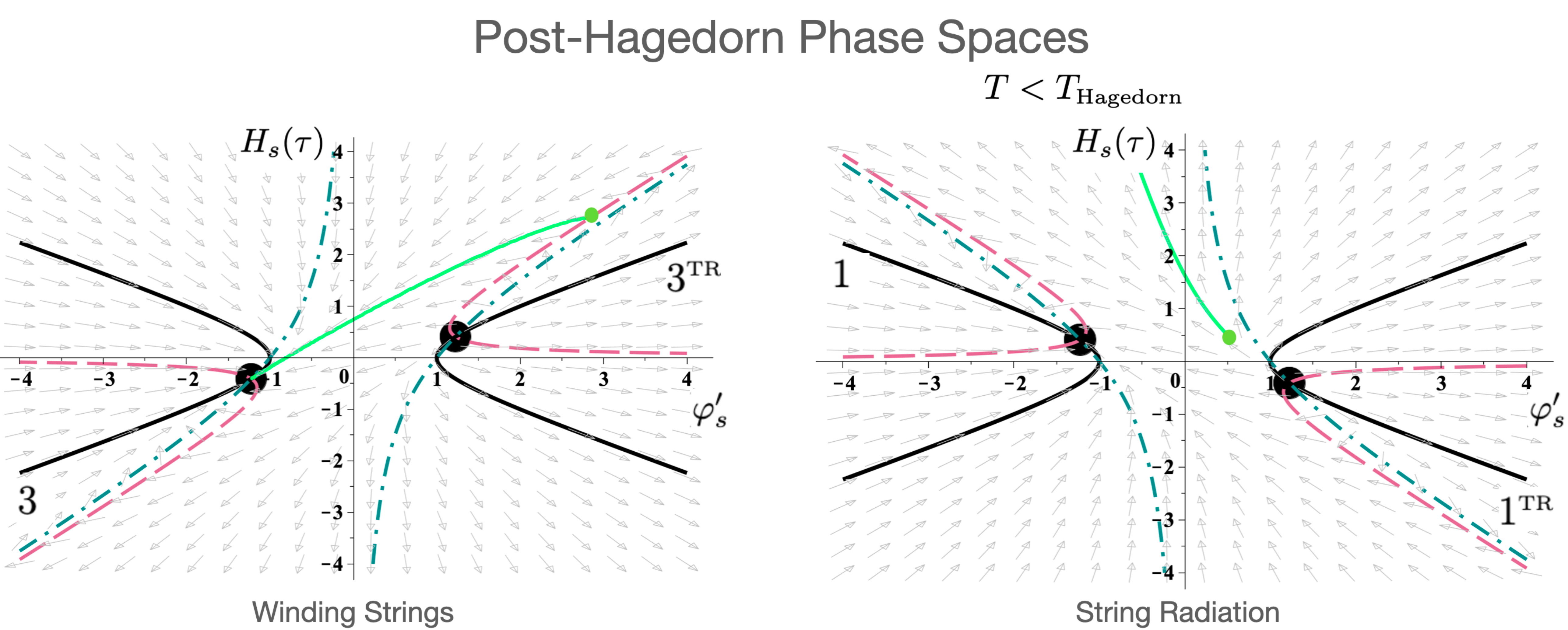}
    \caption{\smaller The phase space ($\varphi_s^\prime,H_s$) of Hagedorn cosmologies below the critical temperature. The black curves enforce the Hamiltonian constraint. The green trajectory is an example of an (unphysical) observer.  The blue (dot-dash) trajectories are lines of constant $\varphi_s^\prime$, whereas pink (dashed) trajectories correspond to constant $H_s(\tau)$.  Their intersection occurs at the causal fixed points $\left(\pm \sqrt{3 /2},\pm     \sqrt{1 /6}  \right)$ and $\left(\mp \sqrt{3 /2},\pm     \sqrt{1 /6}  \right)$ corresponding to winding strings and radiation, respectively. The labeled lines corresponding to the solutions in Table \ref{RadiationTable}. They represent non-bouncing cosmologies in the phase space.}
\label{SGCfig2} 
\end{figure}  
\begin{table}[h]
\begin{center}\begin{tabular}{|c|c|c|c|}
\hline  Region &  Branch   & Expansion & Shifted Dilaton  \\
\hline
1 & $(-)$ &$H_s>0$ & $\vpd_s<0$ \\
$1^{\mbox{\tiny TR}} $& $(+)$ & $H_s<0$ & $\vpd_s>0$ \\
3 & $(-)$& $H_s<0$ & $\vpd_s<0$  \\
$3^{\mbox{\tiny TR}} $ & $(+)$ & $H_s>0$ & $\vpd_s>0$ \\
\hline \end{tabular} 
\caption{\label{RadiationTable} The four regions of the {\bf Post}-Hagedorn phase $(T\lesssim T_H)$ for dynamical solutions that do {\bf not} have a bounce. The first two regions (1 and $1^{\mbox{\tiny TR}} $) correspond to string radiation with $1^{\mbox{\tiny TR}} $ the time reversal. 
Whereas the last columns correspond the winding model duals (3 and $3^{\mbox{\tiny TR}} $).
We have respected the conventions of \cite{Kaloper:2007pw}. 
}
\end{center}
\end{table}

 
To analyze the phase space of the sub-Hagedorn regime we
again use the time redefinition \eqref{tautime} with $\gamma=\pm 1/3$ for $(+)$ radiation / $(-)$winding strings. 
Then the equations are given by  
\eqref{tautimeeoms} with the relevant values of $\gamma$.
We again introduce $l(\tau)=\lp_s$ and $f(t)=\vp_s$ and the equations can be taken as a first order system with fixed points given 
by \eqref{tautimefix}.

The phase spaces are given in Figure \ref{SGCfig2}. There we plot the string frame Hubble parameter in the conformal time $\tau$ versus the derivative of the shifted dilaton. All physically relevant trajectories are restricted by the Hamiltonian constraint to begin and end on the black (co-dimension one) hyperbolae and any branch change would violate the conservation of energy. The green trajectory provides an example of an (unphysical) observer that respects the phase space flow but does not satisfy conservation of energy.  The blue (dot-dash) trajectories are lines of constant (in conformal time) $\varphi_s^\prime$, whereas pink (dashed) trajectories correspond to constant (in conformal time) $H_s(\tau)$.  Their intersection occurs at the causal (hyperbolic) fixed points $\left(\pm \sqrt{3 /2},\pm     \sqrt{1 /6}  \right)$ and $\left(\mp \sqrt{3 /2},\pm     \sqrt{1 /6}  \right)$ corresponding to the phase space flow of winding strings and radiation, respectively. 
We note that the mirror symmetry of the two plots is an example of T-duality in the theory.

\section{Exit from lingering and the Null Energy Condition \label{NECit}}
In this section we review the importance of the egg function\footnote{We would like to emphasize that no chickens were harmed in performing this research, however our eggs are not free range due to the constraints of the NEC.}
in describing violations of the NEC and the viability of models. The importance of this constraint is that NEC violations imply fine-tuning in cosmological models. Inflation does not have this type of fine-tuning, but does have a past singularity (geodesic incompleteness) \cite{Kinney:2021imp,Kinney:2023urn} which, again, is what we are trying to address in this paper. We discuss both the classical model with spatial curvature (Section \ref{two}) and the Hagedorn motivated model (Section \ref{stringcosmo}).

Considering whether a cosmological model violates the NEC is important for determining its viability and predicability. 
NEC violations signal the presence of cosmological singularities and a breakdown of General Relativity \cite{Hawking:1973uf}.
In the context of the evolution of cosmological perturbations NEC violations are not detrimental, a priori, but imply the presence of fine-tuning of initial conditions especially when considering causality of observers \cite{Adams:2006sv} and the evolution of cosmological perturbations. 
NEC violations do not have to be catastrophic, if localized in space-time and in a controlled manner respecting the limited range of validity of the effective field theory \cite{Creminelli:2006xe}.
However, for a complete picture of the universe we will need to assume that the NEC is respected globally.

In the absence of a complete theory of quantum gravity we are forced to assume what is to be known as `reasonable' properties of matter and energy.
A conservative assumption is that sources should lead to positive or zero curvature so that geodesics of the space-time converge along a
null vector $n^\mu$ (matter / energy is always attractive) \cite{Hawking:1973uf}.
This null convergence condition requires $R_{\mu \nu} n^{\mu} n^{\nu} \geq 0$, using Einstein's equation this implies
\be
T_{\mu \nu} n^\mu n^\nu \geq 0. \label{NEC}
\ee
For the FLRW type universes we are considering here, this implies $\rho +p \geq 0$.

\begin{figure} 
    \centering
    
    \begin{subfigure}[b]{.5\textwidth}
         \centering
         \includegraphics[width=\textwidth]{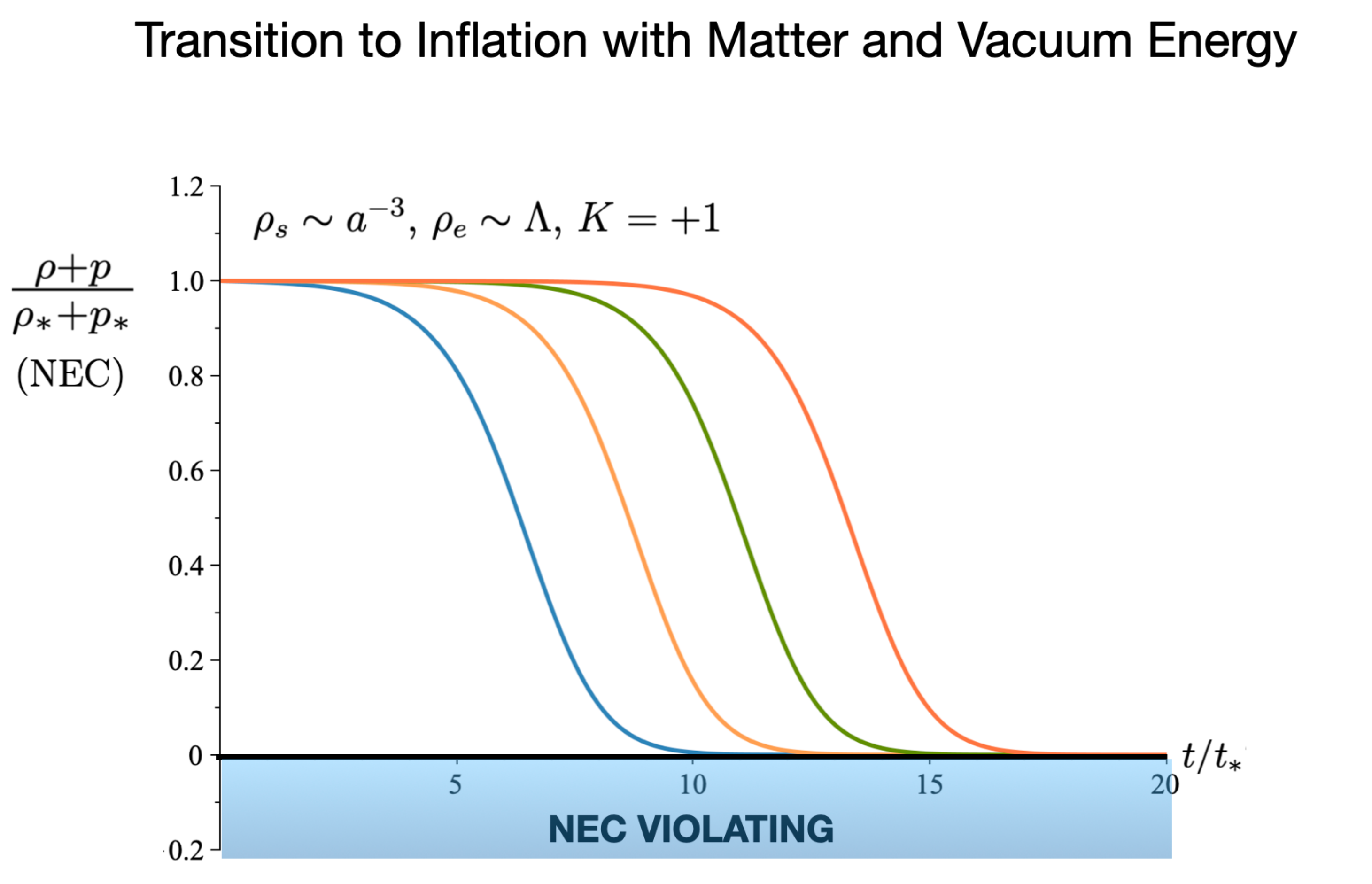}
    \end{subfigure}%
    \begin{subfigure}[b]{.5\textwidth}
         \centering
         \includegraphics[width=\textwidth]{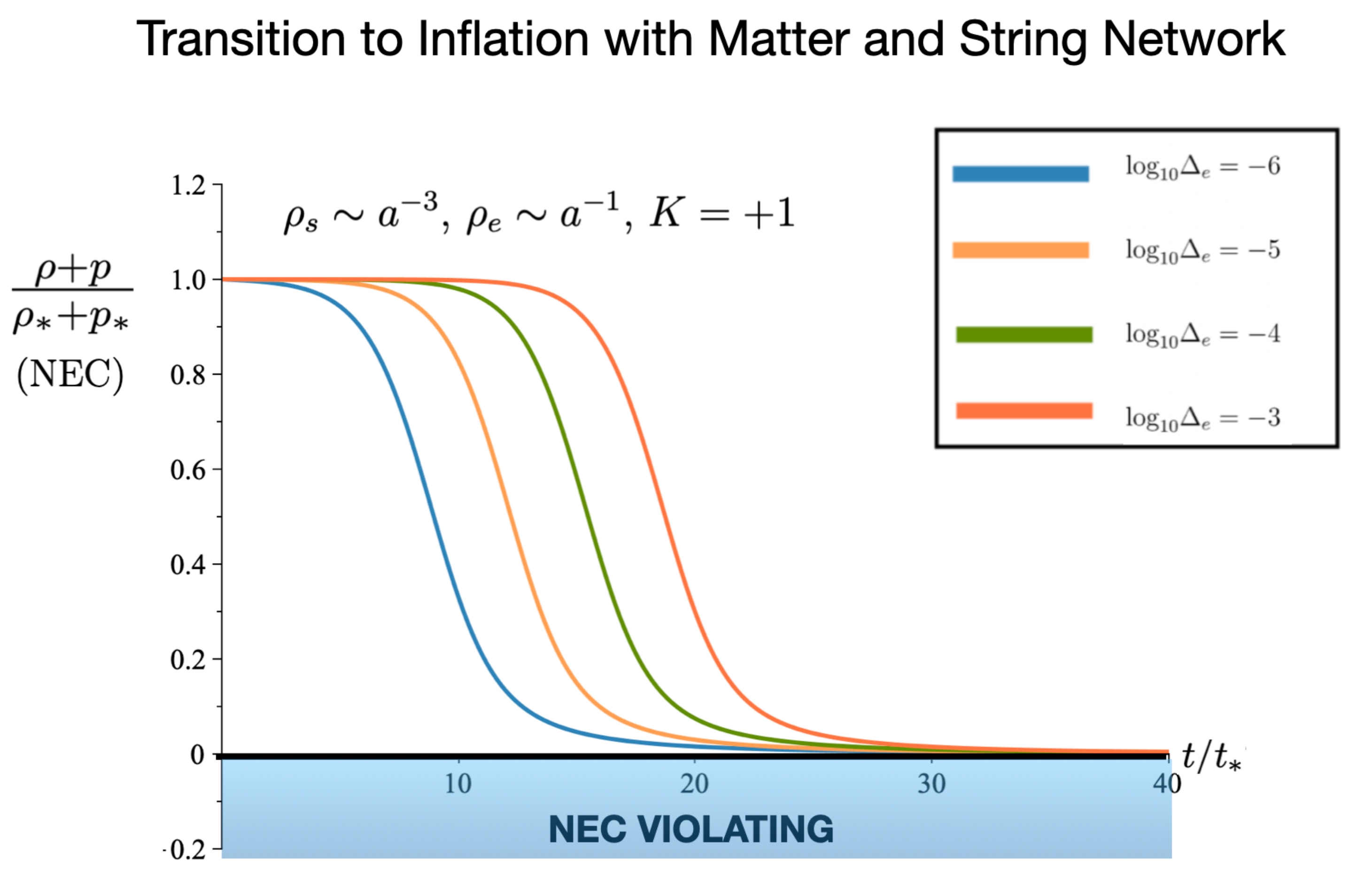}
    \end{subfigure}
    \caption{ \label{fig:NEC} The NEC condition \eqref{NEC} recalling the importance of spatial curvature in \eqref{NECeom}. Above (left) is the NEC as a function of scaled time for a universe with positive spatial curvature, a standard matter component, and positive vacuum energy as it evolves to the inflationary phase. The initial conditions are set by the lingering conditions \eqref{pl4}
    and allowing a perturbation of $\Delta_e$ as discussed in Section \ref{three}.   
    The right plot is the same configuration, however instead of a cosmological constant we consider a string network with equation of state $w_e = -1/3$. 
    We see that the NEC is never violated in both models.}
\end{figure}

In our analysis thus far, we have respected the NEC as we required our second sector to respect $p_e \geq -\rho_e/3$ as discussed in Section \ref{loitloit}.
In particular, the lingering conditions were 
\begin{align} \label{pl4}
\frac{\overline{\rho}_s}{a^m} + \frac{\overline{\rho}_e}{a^n} - \frac{3}{\kappa^2 a^2} = 0,  \\
(m - 2) \frac{\overline{\rho}_s}{a^m} + (n - 2) \frac{\overline{\rho}_e}{a^n} = 0, \nonumber
\end{align}
and these preserve the NEC at all times but also lead to interesting dynamics due to the presence of spatial curvature. 
Indeed, as can be seen from Figure \ref{fig:NEC} the spatial curvature term leads to a saturation of the NEC as can also be deduced from the equations above. 
\\

Our classical analysis of Section \ref{two} implies that the lingering phase can:
\bi
\item Be realized in a universe with positive spatial curvature while preserving the NEC,
\item Can exit to a period of inflation without violating the NEC.
\ei
We now turn to the analysis in Section \ref{stringcosmo} and review the challenges for a String Gas approach as discussed in \cite{Kaloper:2007pw}.

\subsection{Branch change and the egg function}

We reconsider the equations (\ref{eoms_string}), now allowing for additional contributions that could arise from other stringy matter, $\alpha^\prime$ or $g_s$ corrections, spatial curvature, etc. which may be non-trivially coupled to the dilaton. These can be included in the effective Lagrangian ${\cal L}_m={\cal L}_m ( \phi, g_{\mu \nu}, \ldots )$ and in order to preserve spatial isotropy these sources must be of the perfect fluid form\footnote{Although we consider the isotropic case here, given our assumption of spatial homogeneity the presence of anisotropy will not change our conclusions.  As discussed in e.g. \cite{Brustein:1997ny}, 
we could equivalently work in the $4D$ effective theory and then the anisotropies would simply appear as sources in the $4D$ effective `matter' lagrangian.
For our arguments to follow, it is enough to note that such sources (anisotropies) will not violate the Null Energy Condition.}, namely $T_{\mu \nu}=diag \left( \rho_s, p_s, \ldots, p_s \right)$. We rewrite equations (\ref{eoms_string})
\bea 
\dot{\varphi}_s&=& \pm \sqrt{3 \dot{\lambda}_s^2  + e^{\phi_s} \rho_s}, \label{ve1} \\
\dot{H}_s&=&\pm\dot{\lambda}_s \sqrt{3 \dot{\lambda}_s^2+e^{\phi_s} \rho_s} + \frac{1}{2} e^{\phi_s} \left( p_s + \Delta_\phi {\cal L}_m \right), \label{ve2} \\
\dot{\rho}_s&=&-3 \ld_s \left( \rho_s + p_s \right)- \pd  \Delta_\phi {\cal L}_m, \label{ve3} 
\eea
where $ 2 \sqrt{-g} \, \Delta_\phi {\cal L} \equiv \delta {\cal L}_m / \delta \phi_s$ results from the possible coupling of sources to the dilaton. Combining the above we find an equation that will prove useful below,
\be \label{helpful1}
\vdd_s- 3 \lds_s=\half e^{\phi_s} \left( \rho_s - \Delta_\phi {\cal L}_m \right).
\ee

From (\ref{ve1}) we immediately see that in order for a branch change to occur the so-called egg function 
\be  \label{eggfunction}
e=3 \dot{\lambda}_s^2  + e^{\phi_s} \rho_s,
\ee
must vanish, implying that $\rho_s<0$ is required. Hence, the sign of $\vpd_s$ acts as a kind of topologically conserved charge, in the sense that for positive energy solutions a branch change cannot occur classically.

One might optimistically hope that the addition of string corrections, non-trivial coupling to the dilaton, and/or other contributions to the effective action that appear in ${\cal L}_m$ might allow for $\rho_s<0$ and a successful exit from the Hagedorn phase.
Entertaining such a possibility let us continue assuming that $\rho_s<0$ is achieved and ask whether this is sufficient to achieve an exit from the Hagedorn phase.

A useful relation can be found by using $\vpd_s= \pm \sqrt{e}$ and (\ref{helpful1}),
\bea \label{helpful2}
\pm\frac{d}{dt} \left( \sqrt{e} \right) &=&  3 \lds + \half e^{\phi_s} \left( \rho_s - \dlm \right),\label{help3} \\
 &=&  e + \half e^{\phi_s} \left| \rho_s \right| - \half e^{\phi_s} \dlm, 
\eea
where in the last step we have used the definition of the egg function (\ref{eggfunction}) and the fact that $\rho_s<0$ as we approach the egg ($e=0$). {$(\pm)$ refers to the branch on which this evolves.}

First let us consider the case when the sources are trivially coupled to the dilaton in the string frame, so that $\dlm=0$.
We are interested in the possibility that we initially approach the egg from the Hagedorn $(+)$ branch and then make a transition to the RDU $(-)$ branch.
Because we have seen that requiring analyticity implies $e \geq 0$ (classically allowed region), it follows that $\frac{d}{dt} (\sqrt{e}) < 0$ in order for the transition to occur. 
In the case of trivial dilaton coupling (i.e. $\dlm$=0) we find 
\be 
\frac{d}{dt} \left( \sqrt{e} \right)= e + \half e^{\phi_s} \left| \rho_s \right| >0,
\ee
where we have used $\rho_s<0$ before the transition.  We immediately see that a branch change cannot occur, since the egg is repulsive, i.e. the point $e=0$ is a repeller in the phase space and trajectories will not lead to successful transitions.

Let us now consider the case of non-trivial coupling to the dilaton.  From $(+)$ branch solutions of
(\ref{helpful2}), we might naively expect that a branch change could occur if contributions from $\dlm$ are negative enough to change the overall sign.
We will now show that is not the case, and will prove a no-go theorem for the Hagedorn exit.

\subsection{A no-go theorem for Hagedorn exit}
Using (\ref{ve2}), we eliminate $\dlm$ from {(\ref{help3}) giving}
\be
{\pm}\frac{d}{dt} \left( \sqrt{e} \right) = -\ldd_s+\ld_s \pd_s + \half e^{\phi_s} \left({ \rho_s } + p_s \right).
\ee
We integrate the above equation from the moment of the branch change $t_h$ when $e=0$, to the moment of escape $t_e$ when $\rho_s=0$ ($\rho_s>0$ thereafter) and $e=3\lds_s$. {For evolution along the $(-)$ branch, this gives}
\be
\int_{t_h}^{t_e} dt \left[ \frac{d}{dt} \left( \textcolor{blue}{-}\sqrt{e} + \ld_s \right) -\ld_s \pd_s \right] = \half  \int_{t_h}^{t_e} dt  \; e^{\phi} \left( \rho_s + p_s \right).
\ee
Evaluating the integral, we find
\bea
 \left(\sqrt{3}-1\right) \ld_s(t_e) + \ld_s(t_h) + {\cal A}  &=& -\half  \int_{t_h}^{t_e} dt  \; e^{\phi} \left( { \rho_s} + p_s \right),
\eea
where ${\cal A}=\int \ld_s d\phi >0$ is the positive definite volume in the phase space.
The left hand side is always positive, since $\ld_s >0$ on both branches.  This implies that an exit from the Hagedorn phase requires $ \rho_s +p_s <0$, {\it i.e.}, a violation of the NEC. One way to avoid this constraint is to introduce spatial curvature as we discussed in the `classical situation'. In previous results of String Gas Cosmology spatial curvature was ignored because inflation did not take place (it was posed as an alternative to inflation). But considering a lingering phase leading to inflation means that this is no longer an issue and could resolve the problems with NEC violation that we have discussed above. 

In the string based models we are discussing here, they would simply start in the lingering phase and then evolve into dilaton led inflation. This can be seen from Figure \ref{SGCfig}. The phase space flow leads to a repeller (hyperbolic) fixed point, which then leads to inflation -- there is no branch change. There are many open questions we must address. How long does loitering last -- how does inflation end? These are challenging questions we must address in future research. The promising aspect of this work is it suggests a new way to look at the beginning of inflation.

\section{Conclusions}
{\bf What we are not proposing.}
In the models we have presented above we considered solutions where the cosmological evolution moves toward a state of inflationary cosmology (quasi-dS space-time). That is, thus far we are relying on inflation to address the horizon, relic, flatness, etc... problems \cite{Guth:1980zm,Linde:1981mu,Albrecht:1982wi}. 
{\it We are not proposing a complete alternative to inflation}.
Instead, we are proposing an alternative way to look at the beginning of inflation -- the universe lingered. 
For minimal inflation there is an interesting possibility of small, but non-zero spatial curvature which would be clarified by CMB-S4 \cite{Abazajian:2016yjj} and other future observations.   The motivation for this approach was to provide a new way to address the past cosmological singularity of inflation and the initial state of the universe. In future work we will consider the effect (if any) on the cosmological perturbations in the model. 

We discussed a more provocative motivation in Section \ref{stringcosmo}.
In string theory approaches to early universe cosmology, string dynamics can lead to a Hagedorn phase which cosmologically implies a lingering phase. As discussed above, the lingering phase does not demand the presence of spatial curvature and can instead arise from the dynamics of a string gas and an evolving dilaton. Thus far, the only known solution that would not violate the NEC naturally leads to inflation, as shown in  \cite{Kaloper:2007pw}.
Our analysis here has used spatial curvature as a proxy for the string dynamics discussed above which will be considered in future work.

We have seen that the initial state of the universe can differ from that of a dS space-time and then evolve into an inflationary state without violating the Null Energy Condition. This suggests a new class of models for addressing the problem of initial conditions for inflation and related issues.
In particular, the lingering phase and transition are motivated by the Hagedorn phase of string theory when applied to cosmology.
Although this shares many similarities with models of String Gas Cosmology, it differs in that the transition from the initial state is to {\it inflation} and this transition does not violate the Null Energy Condition. The latter is crucial for preserving the predictiveness and causality of the theory, particularly in calculating the initial state of the inflationary cosmological perturbations. This is work in progress.

Points of enhanced symmetry can act as dynamic attractors like the lingering point we have discussed above
\cite{Banks:1994sg,Kofman:2004yc,Watson:2004aq,Cremonini:2006sx}.
If all moduli were stabilized in this way the theory would be at strong coupling.
New analytic techniques are needed to explore such a regime, however the situation is not that different from QCD. In the case of an enhanced symmetry point associated with the radii of the spatial dimensions one could view this as a confining gauge theory.
That is, at the duality point there are additional light degrees of freedom and the associated gauge group is promoted to an $SU(2)$ gauge theory.
It was shown in \cite{Watson:2004aq} that considering the backreaction of these particles on the dynamics would lead to a confining potential or `moduli trapping'. That is, as the dimensions evolve away from the fixed point the gauge bosons become Higgsed which is not energetically favored (see also \cite{Cremonini:2006sx,Greene:2007sa}). 
We leave exploring these ideas and the physics of the Hagedorn / lingering phase to future publications.

\section*{Acknowledgements}
We thank Simon Catterall, Nemanja Kaloper, Hiroshi Ooguri, Kenny Ratliff, Gary Shiu, Eva Silverstein, Kuver Sinha, and Cumrun Vafa for useful conversations.  We especially thank Will Kinney for useful discussions and hospitality. S.W. thanks KITP Santa Barbara and the Simons Center for hospitality and financial support. This research was supported in part by DOE grant DE-FG02-85ER40237. 

\appendix
\section{Einstein Frame Hubble parameter and scale factor asymptotics \label{app1}}
One can move between the Einstein and String frames as (see e.g. \cite{Battefeld:2005av})
\be
\label{S2E}
{g^e}_{\mu\nu} = e^{-2 \phi_s / (N - 1)} {g^s}_{\mu\nu}, \quad \phi_e = \sqrt{\frac{2}{N-1}} \phi_s.
\ee
One then finds that
\begin{align}
\ln a_e &\equiv \lambda_e = - \frac{\lambda_s + \varphi_s}{N-1}, \\
H_e &\equiv \frac{d\lambda_e}{d\tau_e} = \frac{d\lambda_e}{d x} \frac{dx}{dt} \frac{dt}{d \tau_e} = \frac{d\lambda_e}{d x} \frac{dx}{dt} \frac{dt}{d\tau_e}.
\end{align}
Using \eqref{S2E} and the definitions of $E$ and $\varphi_s$, the expression for the Einstein frame Hubble parameter, in terms of the variable $x$, becomes
\be
\begin{split}
H_e &= \frac{d\lambda_e}{d x} E_0 e^{-\gamma N \lambda_s} e^{\phi_s / (N-1)} \\
&= \frac{d\lambda_e}{d x} E_0 \exp{\frac{1}{N-1}(\varphi_s + N \lambda_s (1 - \gamma (N-1) )}.
\end{split}
\ee
For convenience, we define $\tilde{\lambda}_e \equiv (N-1) \lambda_e$, $\tilde{H}_e \equiv \frac{d \tilde{\lambda}_e}{d\tau_e}$, and
\begin{align*}
\Theta(A,B) &= A \lambda_s + B \varphi_s \\
&= A \lambda_{s,0} + B \varphi_{s,0} + \ln(x(x-x_\ast)) \left(\frac{\gamma A}{\alpha} - \frac{B}{\alpha}\right) + \ln(1 - \frac{x_\ast}{x}) \left(\frac{A}{\alpha \sqrt{N}} - \frac{\gamma B \sqrt{N}}{\alpha}\right) \\
&= X(A,B) + l_1(x,x_\ast) Y(A,B) + l_2(x,x_\ast) Z(A,B).
\end{align*}
We note $\tilde{H}_e = H_e$
and $\Theta(\epsilon A, \epsilon B) = \epsilon \Theta(A,B).$
Given this we have
\begin{align}
\tilde{\lambda}_e &= - \Theta(1,1), \\
\tilde{H}_e &= - \frac{d \Theta(1,1)}{dx} E_0 \exp(\frac{1}{\tilde{N}}\Theta(1,(\tilde{N} + 1) (1 - \gamma \tilde{N}))),
\end{align}
where $\tilde{N} = N - 1$.

Our goal is to find the Hubble parameter power-law relation with the scale factor, as seen in the Einstein frame. Due to the disconnected nature of the allowed range of $x$ for the solutions in the text, we have four limits of interest. Two are the limits of the behavior in the infinite past of one solution and the infinite future of the same solution on another branch. The second two limits are the approach to $x = 0$ or the transit away from $x \rightarrow x_\ast$.
In the dilaton-dominated Hagedorn phase: $\gamma = 0$ and $\alpha = 1$. One can show that
\be
a = \tilde{a}^{1/\tilde{N}} = e^{\tilde{\lambda}_e / \tilde{N}} = a_0 \left(x (x - x_\ast) \right)^\sigma \left(1 - \frac{x_\ast}{x} \right)^\lambda,
\ee
where $a_0 = \exp(-(\lambda_{s,0} + \varphi_{s,0}) / \tilde{N})$, $\sigma = \tilde{N}^{-1}$, and $\lambda = -\left(\tilde{N} \sqrt{\tilde{N}+1} \right)^{-1}$. A number of further substitutions and expansions also lead to
\be
H_e = H_0 \left(x (x - x_\ast) \right)^{-\gamma} \left(1 - \frac{x_\ast}{x} \right)^\delta \left(x - \Gamma x_\ast\right),
\ee
where we have defined
\begin{align}
H_0 &= 2 E_0 \exp(\frac{1}{\tilde{N}} (\lambda_{s,0} + (\tilde{N} + 1) \varphi_{s,0}) ), \\
\gamma &= \frac{1 + 2 \tilde{N}}{\tilde{N}}, \\
\delta &= \left(\tilde{N} \sqrt{\tilde{N} + 1} \right)^{-1}, \\
\Gamma &= \frac{1 + \sqrt{\tilde{N} + 1} }{\sqrt{\tilde{N} + 1} }.
\end{align}

When $x \rightarrow \pm \infty$, we find that
\be
\frac{a}{a_0} \sim x^{2\sigma}, \quad \frac{H_e}{H_0} \sim x^{1 - 2 \gamma} \Rightarrow \frac{H_e}{H_0} \sim \left( \frac{a}{a_0}\right)^{\frac{1-2\gamma}{2\sigma}} = \left( \frac{a}{a_0}\right)^{\frac{1}{2}(1-3N)}.
\ee

When $x \rightarrow 0$, we find that
\be
\frac{a}{a_0} \sim x^{\sigma - \lambda}, \quad \frac{H_e}{H_0} \sim x^{-(\gamma + \delta)} \Rightarrow \frac{H_e}{H_0} \sim \left( \frac{a}{a_0}\right)^{\frac{\gamma + \delta}{\lambda - \sigma}} = \left( \frac{a}{a_0}\right)^{-(1 - 2 \sqrt{N} + 2N)}.
\ee

When $x \rightarrow x_\ast$, we find that
\be
\frac{a}{a_0} \sim (x - x_\ast)^{\sigma + \lambda}, \quad \frac{H_e}{H_0} \sim (x - x_\ast)^{\delta - \gamma} \Rightarrow \frac{H_e}{H_0} \sim \left( \frac{a}{a_0}\right)^{\frac{\delta - \gamma}{\sigma + \lambda}} = \left( \frac{a}{a_0}\right)^{-(1 + 2 \sqrt{N} + 2N)}.
\ee

For a universe dominated by a perfect fluid with equation of state $w$, standard cosmology implies that $H \sim a^{-\frac{3}{2}(1+w)}$. The effective equations of state for the dilaton dominated universe are
\begin{align}
    x \rightarrow \pm \infty: &\quad w = N - \frac{4}{3}, \\
    x \rightarrow 0 : \quad w = -\frac{1}{3}(1 + 4 \sqrt{N} - 4 N), & \quad \quad x \rightarrow x_\ast : \quad w = \frac{1}{3}(4 N + 4 \sqrt{N} - 1).
\end{align}

\bibliographystyle{apsrev4-1}
\bibliography{paper.bib}

\end{document}